\documentclass[proceedings, preprint]{rmaa}
\usepackage{paralist}
\usepackage{psfrag,color}
\usepackage{flushend}
\usepackage{cuted}
\usepackage{graphicx}	
\usepackage{amsmath}	
\usepackage{xspace}
\usepackage{wrapfig}
\usepackage{dcolumn}
\usepackage{natbib}
\usepackage{float}
\usepackage{caption}
\usepackage{placeins}
\usepackage{rotating}
\usepackage{enumitem}
\usepackage{lastpage}
\usepackage{subcaption}
\usepackage[none]{hyphenat}
\usepackage{threeparttable}
\usepackage{multicol,amsmath}
\usepackage{newtxtext,newtxmath}
\usepackage[table,x11names]{xcolor}
\usepackage[margin=0.78in]{geometry}
\usepackage{multirow,longtable,supertabular}
\usepackage[colorlinks=true,citecolor=blue, linkcolor=blue, urlcolor=blue]{hyperref}

\newcommand{\fermi}{{\em Fermi}\xspace}
\newcommand{\swift}{{\em Swift}\xspace}
\newcommand{\Ep}{E$_{\rm p}$\xspace}
\newcommand{\sw}[1]{\texttt{#1}}
\newcommand{\tninty}{T$_{\rm 90}$\xspace}
\newcommand{\tzero}{T$_{\rm 0}$\xspace}
\newcommand{\na}{New Astronomy}
\SetYear{2023}
\SetConfTitle{7th Workshop on Robotic Autonomous Observatories}

\title{Prompt and afterglow analysis of the \fermi-LAT detected GRB 230812B}
\date{15 January 2024} 

\author{
Amit K. Ror \altaffilmark{1,2}, 
S. B. Pandey \altaffilmark{1}, 
A. Aryan \altaffilmark{1,3},
Sudhir Kumar \altaffilmark{2},
and A. J. Castro-Tirado \altaffilmark{4}}

\altaffiltext{1}{Aryabhatta Research Institute of Observational Sciences (ARIES), Manora Peak, Nainital-263002, India.}

\altaffiltext{2}{Department of Applied Physics/Physics, Mahatma Jyotiba Phule Rohilkhand University, Bareilly-243006, India.}

\altaffiltext{3}{Graduate Institute of Astronomy, National Central University, 300 Jhongda Road, 32001 Jhongli, Taiwan}

\altaffiltext{4}{Instituto de Astrof\'isica de Andaluc\'ia (IAA-CSIC), Glorieta de la Astronom\'ia s/n, E-18008, Granada, Spain}

\shortauthor{Amit K. Ror et al.}
\shorttitle{GRB 230812B}

\resumen{La emisi\'on temprana de GRB 230812B se destaca como uno de los eventos m\'as luminosos observados por instrumentos del sat\'elite Fermi tales como GBM y LAT. El an\'alisis espectral de emisi\'on temprana (tanto integrado en el tiempo como  en sus diversas fases temporales) de esta explosi\'on respalda una componente t\'ermica adicional y otra no t\'ermica, lo que indica una composici\'on h\'ibrida del jet (chorro). Los par\'ametros espectrales $\alpha$, E$_p$ y kT del modelo espectral de Band junto a cuerpo negro son los que proporcionan un mejor ajuste. Adem\'as, las emisiones de la postluminiscencia a energ\'ias m\'as bajas son consistentes con la emisi\'on de sincrotr\'on del choque externo que se va propagando hacia adelante barriendo el medio del medio interestelar. El instrumento LAT detect\'o una emisi\'on de muy alta energ\'ia (VHE) que se desv\'ia de los esperado del mecanismo de emisi\'on sincrotr\'on, posiblemente originada por el aumento del factor de  Lorentz de los fotones de emisi\'on temprana motivado por electrones acelerados en el choque externo a trav\'es de mecanismos de emisi\'on de Compton inverso (IC) o auto sincrotr\'on Compton (SSC). La comparaci\'on de las propiedades de emisi\'on temprana y de la posterior de este estallido revel\'o que, a diferencia de la emisi\'on simult\'anea (gamma) y m\'as brillante, la postluminiscencia posterior de GRB 230812B es m\'as d\'ebil que la de los otros estallidos de rayos gamma tambi\'en brillantes y asociados con una supernova (GRB 130427A y GRB 171010A) con desplazamientos al rojo similares.}

\abstract{Prompt emission of GRB 230812B stands out as one of the most luminous events observed by both the \fermi-GBM and LAT. Prompt emission spectral analysis (both time-integrated and resolved) of this burst supports an additional thermal component {together with a non-thermal}, indicating the hybrid jet composition. {The spectral} parameters $\alpha$, \Ep, and kT of the best-fit \sw{Band+Blackbody} model {show a tacking behaviour with the intensity}. Further, the low energy afterglow emission {is} consistent with the synchrotron emission from the external forward shock in the ISM medium. LAT detected very high energy emission (VHE) deviating from the synchrotron mechanism, possibly originating from the Lorentz boosting of prompt emission photons by accelerated electrons in the external shock via Inverse Compton (IC) or Synchrotron Self Compton (SSC) emission mechanisms. The comparison of the prompt and afterglow emission properties of this burst revealed that, unlike the bright prompt emission, the afterglow of GRB 230812B is fainter than the other SN-detected bright bursts (GRB 130427A and GRB 171010A) at a similar redshift.}

\addkeyword{Gamma ray burst}
\addkeyword{mechanism: Blackbody, Synchrotron, Synchrotron self Compton.}

\begin{document}
\maketitle

\section{Introduction}
Gamma-ray bursts (GRBs) are distant and powerful cosmic events characterized by two phases: A highly variable prompt emission in soft $\gamma$-rays and hard X-rays {with a duration of} (\tninty) seconds (s) to minutes, followed {by a} smoothly decaying afterglow emission spanning a broad temporal (hours-days) and {energy ranges} (radio-TeV {energies; \citealt{2015PhR...561....1K}}). GRBs are traditionally classified based on temporal and spectral properties into short/hard (SGRB) and long/soft {(LGRBs; \citealt{1993ApJ...413L.101K})}. SGRBs (\tninty $<$ 2\,s) {come from} compact binary mergers, and LGRBs (\tninty $>$ 2\,s) {are likely} originating {from a} massive star collapse \citep{1998Natur.395..670G, 2003Natur.423..847H, 2017ApJ...848L..13A}. However, GRB 200826A (\tninty $\sim$ 1.2\,s from {collapsar; \citealt{2021NatAs...5..917A}}), GRB 211211A {(\tninty $\sim$ 50\,s; \citealt{2022Natur.612..228T}}), and {GRB 230307A (\tninty $\sim$ 35\,s; \citealt{2023arXiv230702098L}}) {come from} {mergers} are outlier to the traditional classification. Newly introduced classifications include intermediate bursts and ultra-long GRBs {(ULGRBs, \tninty $>$ 1000\,s; \citealt{1998ApJ...508..314M, 2015ApJ...800...16B, 2024ApJ...971..163R})}, broadening the classification beyond traditional LGRBs and SGRBs.

Over the 50 years of study, the GRB prompt emission remains a subject of ongoing exploration \citep{2015AdAst2015E..22P}. Several theories describe its origin from internal shock or magnetic reconnection in the ultra-relativistic jet. The jet composition can be baryonic or magnetic, and the mechanism of prompt emission from these ultra-relativistic jets (\sw{synchrotron}, non-thermal, thermal, or hybrid) is still an open question \citep{2018NatAs...2...69Z, 2024ApJ...972..166G}. A widely employed \sw{Band} function \citep{1993ApJ...413..281B}, introduced during the {\textit{BATSE}} era, characterizes {the non-thermal} spectral components of GRBs. It incorporates four parameters: (1) normalization constant, (2) low-energy spectral index ($\alpha$ $\sim$ -1), (3) high-energy spectral index ($\beta$ $\sim$ -2), {and (4)} peak energy {(\Ep $\sim$ 300 keV; \citealt{2000ApJS..126...19P})}. However, results of time-resolved spectral analysis from {\citeauthor{2006ApJS..166..298K} \textcolor{blue}{(2006)}} have revealed variations, with a harder $\alpha$ $\sim$ -0.72. Subsequent studies of 1500 time-resolved spectra from 78 bright bursts \citep{2024RAA....24b5006W} revealed {that the} $\alpha$ values from 80\% of spectra crossing the limit (-3/2,-2/3) {are due to} the fast and slow cooling limits from \sw{synchrotron} emission mechanism known as \sw{synchrotron} line of death {\citep{2000ApJS..126...19P}}. Recently, {new models} have been used to fit complicated {spectra} {like the} physical \sw{synchrotron} and \sw{blackbody} to constrain the emission mechanism {\citep{2020NatAs...4..174B}}. Studies have shown that in the case of GRBs, where $\alpha$ crosses the limit (-3/2,-2/3), still the physical \sw{synchrotron} model is found {to best fit} {the spectra} in some cases \citep{2020NatAs...4..174B}. In addition, {the study of the spectral} parameter evolution during the prompt emission serves as a valuable tool to constrain emission mechanisms. The peak energy \Ep shows four types of evolution: flux tracking, hard to soft, soft to hard, and a chaotic \citep{1983Natur.306..451G, 1986ApJ...301..213N}. Time-resolved spectral analysis of bright GRBs from \fermi-GBM revealed that the \Ep obtained for most of the cases shows flux tracking, but hard to soft evolution is common in single pulse GRBs \citep{2024RAA....24b5006W, 2021ApJS..254...35L}. The evolution of $\alpha$ is, on the other hand, less predictable. However, recently {\citet{2021MNRAS.505.4086G, 2022MNRAS.511.1694G, 2023arXiv231216265G}; \citet{2023ApJ...942...34R} and \citet{2024BSRSL..93..709R}} determined the flux tracking evolution of $\alpha$ in GRB 140201A and GRB 201216C, respectively.

GRB afterglows from the cooling population of accelerated electrons from external {shocks} generally decay following a simple or broken PL. The late-afterglow light curve (hereafter LC) of some nearby GRBs exhibits a distinct bump, indicating the emergence of the underlying supernova {(SN; \citealt{1998Natur.395..670G, 2003Natur.423..847H})}. Detailed spectroscopic examinations of these underlying SNe reveal that they are predominantly hydrogen-deficient broad-line type Ic-BL, often categorized as striped envelope supernovae (SESNe), for more detail about SESNe please refer to \citet{2021MNRAS.507.1229P}. Notably, SN-associated GRBs display a luminosity of three to five orders of magnitude less than typical LGRBs, {so detectable} only {in the} close proximity up to a redshift of $\sim$ 1. The energy source of GRB-associated SNe is believed to be thermal heating from radioactive Nickel trapped within the ejecta. Recently, the magnetar model has emerged as a compelling framework to explain the power source for GRB-associated SNe \citep{2022NewA...9701889K}. Another plausible energy source for GRBs associated with SNe involves the interaction between the ejected material and the circumstellar medium surrounding the burst. In the case of GRB 230812B, \citet{2024ApJ...960L..18S} {used} the radioactive heating model to explain the power source of SN 2023pel {to constrain} the radioactive Nickel mass M$_{Ni}$ = 0.38 $\pm$ 0.01 M$_{\odot}$ and the total ejecta mass of 1.0 $\pm$ 0.6 M$_{\odot}$. Recently, 
\citet{wang2024GRB230812B} demonstrated that GRB 230812B is one of the shortest long-duration GRBs (T\(_{90}\) $\sim$ 3\,s) with a collapsar origin. This event is analogous to GRB 200826A \citep{2021NatAs...5..917A}, where a longer jet-bore time through the stellar envelope was proposed to account for the shorter burst duration.

Until 2018, LAT was the only very high-energy instrument that successfully detected GeV emissions {from} more than 100 GRBs \citep{2019ApJ...879L..26F}. LAT detected the maximum energy photon from GRB 130427A {with an energy of} $\sim$ 95 GeV \citep{2014Sci...343...42A}. After 2018, six long GRBs (GRB 180720B, GRB 190114C, 190829A, GRB 201015A, GRB 201216C, and GRB 221009A) were detected as having TeV emissions associated with their prompt and afterglow emissions \citep{2021RMxAC..53..113G, 2023ApJ...942...34R, 2024ApJ...971..163R}. A single synchrotron component could not explain the double bump observed in the broadband SED of VHE-detected GRBs, and an additional Synchrotron Self Compton (SSC) or inverse Compton (IC) component is required \citep{2019ApJ...879L..26F, 2019Natur.575..459M, 2021ApJ...918...12F, 2021Sci...372.1081H, 2023ApJ...942...34R}. In our work, we have tried to constrain the possible origin of GeV detection associated with the GRB 230812B.

The content of the paper is as follows: {\S2 is devoted to multi-band} observation and analysis of GRB 230812B, \S3 {to results}, \S4 {to discussion}, and \S5 {to summary and conclusion}. In this paper, the chosen cosmological parameters are the Hubble constant H$_{\rm 0}$ = 71 km s$^{-1}$ Mpc$^{-1}$ and the density parameters $\Omega_{\rm \Lambda}$ = 0.73 and $\Omega_{\rm m}$ = 0.27. Uncertainties are given {at} 1$\sigma$ level if not stated otherwise. The representation of afterglow flux follows the convention expressed by F($\nu$, t) = t$^{-\alpha}$$\nu^{-\beta}$.

\section{Multi-wavelength Observation and analysis}
\subsection{Prompt emission observation and analysis}
\begin{figure}[!t]
\centering
\includegraphics[width=\columnwidth]{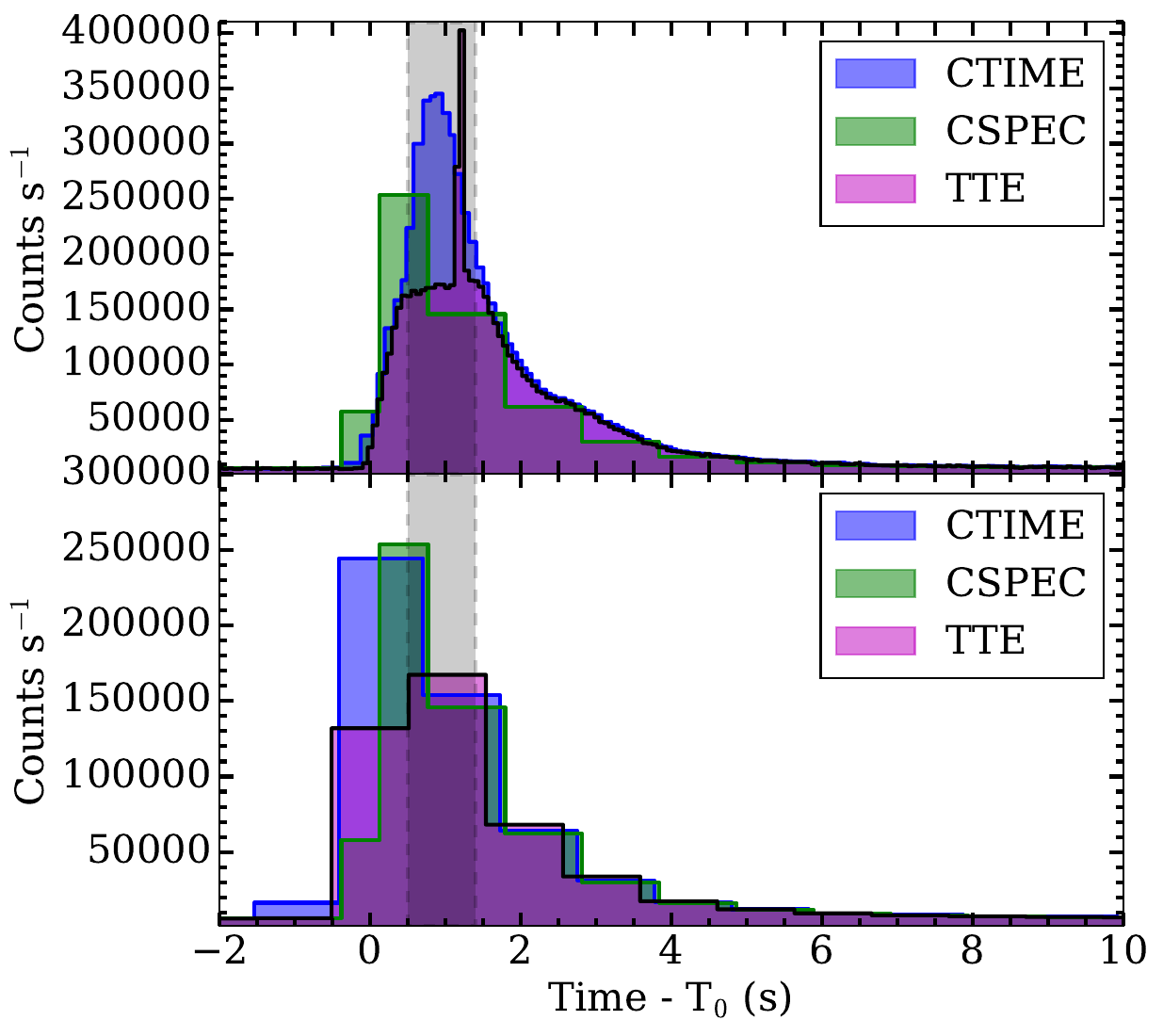}
\caption{Represents the \fermi-GBM LCs of GRB 230812B, plotted from the CSPEC (green), CTIME (blue), and TTE (magenta) data files. The Upper panel shows the 1s binned LC of CSPEC and 64 ms of CTIME and TTE observation. Similarly, the lower panel shows the LC binned with 1s for all CSPEC, CTIME, and TTE. The vertical gray region corresponds to the pulse pileup, which we removed from our data analysis.}
\label{fig:lc}
\end{figure}

\begin{figure}[!t]
\centering
\includegraphics[width=\columnwidth]{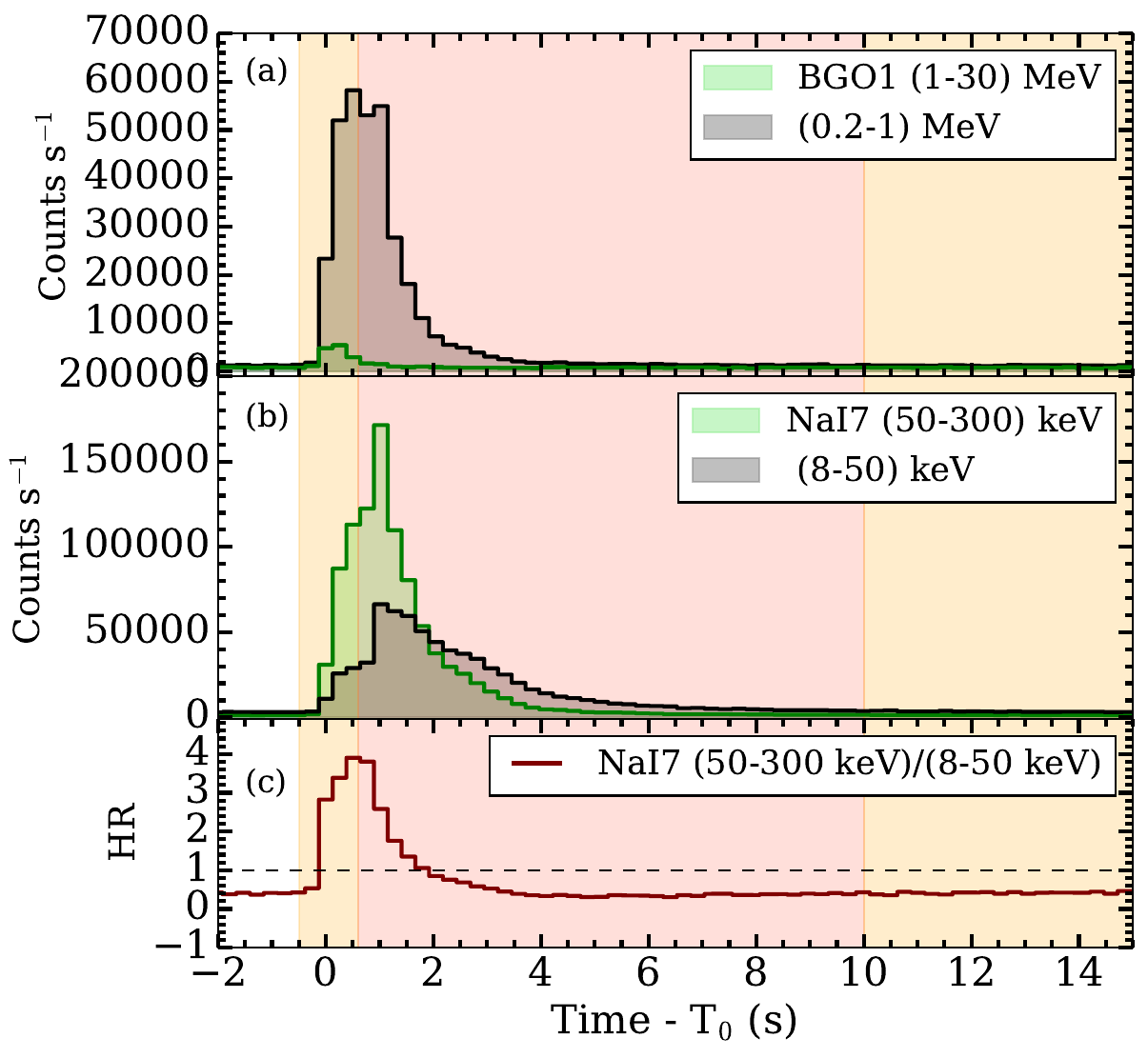}
\caption{The multi-wavelength prompt emission LC of GRB 230812B is plotted from the GBM TTE observation. (a) GBM-BGO observation plotted in the 0.2-1 MeV (gray) and 1-30 MeV (green). (b) Similarly, the NaI LCs in the energy ranges 50-300 keV and 8-50 keV are plotted. (c) Evolution of hardness ratio (HR, number of photons detected in 50-300 keV / number of photons in the 8-50 keV) from the combined NaI-3, NaI-6, and NaI-7 scintillation detectors of \fermi-GBM.}
\label{fig:mclc}
\end{figure}

On August 12, 2023, at 18:58:12.05 UT (hereafter \tzero), the Gamma-Ray Burst Monitor (GBM, \citealt{2009ApJ...702..791M}) on board Fermi gamma-ray space telescope (\fermi) discovered and localised GRB 230812B \citep{2023GCN.34386....1F, 2023GCN.34387....1L}. During the prompt emission phase, the initial 3.3 $\pm$ 0.1\,s exhibited 95\% of the burst's {fluence} in $\gamma$-rays, hence considered the \tninty duration. However, a softer tail persisted for approx 10 s post-detection \citep{2023GCN.34387....1L, 2023GCN.34391....1R}. Prompt {emission's} preliminary analysis revealed that the time-integrated spectra within the 0-32\,s time-interval were well-described by the band function, yielding parameters of \Ep = 273 $\pm$ 3\,keV, $\alpha$ = -0.80 $\pm$ 0.01, and $\beta$ = -2.47 $\pm$ 0.02. The fluence recorded during this period was 2.5201 $\pm$ 0.0002 $\times$ 10$^{-4}$ erg cm$^{-2}$ {by \citet{2023GCN.34391....1R}}. In addition to \fermi-GBM, \textit{GECAM}-C \citep{2023GCN.34401....1X}, {\textit{Konus}}-Wind \citep{2023GCN.34403....1F}, and {\textit{AGILE}}/MCAL \citep{2023GCN.34402....1C} also detected the intense prompt emission from GRB 230812B.

GBM observations for this burst were acquired from the official web page of \fermi Science Support center\footnote{\url{https://fermi.gsfc.nasa.gov/ssc/data/access/gbm/}}. Before analyzing the \fermi-GBM data, we carefully check the observed LC from the three types of spectral file (CTIME, CSPEC, and TTE) using the GBM-Tool python package \citep{gbmdatatool}. In the 64 ms bin LC, we found a glitch around [1.12-1.312] s in the TTE data. A thin magenta spike in the TTE LC at around 1.2 s (see upper panel of Figure \ref{fig:lc}) corresponds to the artificially generated spike mentioned by \citeauthor{2023GCN.34694....1R} \textcolor{blue}{(2023)}. This bright spike is not present in the CSPEC and CTIME data files, which might be due to their coarse temporal or spectral resolution. However, the LC plotted with 1.024 s resolution from the TTE, CSPEC, and CTIME data files showed similar results without any discrepancy (see bottom panel of Figure \ref{fig:lc}). TTE data in the interval [1.12-1.312] s was lost due to its {band-width} limit, and also, there is observed pileup in all types of data files during the {time-interval} [0.5-1.4] s \citep{2023GCN.34694....1R}. We have not utilized the data corresponding to this time range during the time-resolution spectral fitting of GBM data. As recommended by the \fermi-GBM team, we have used NaI-3,6,7 and BGO-0 detectors for the GBM data analysis. All other detectors are blocked by different parts of the spacecraft and cannot be considered reliable \citep{2023GCN.34694....1R}.

We closely followed the methodology described in \citet{2023ApJ...942...34R}, for the prompt emission spectral analysis. The time-integrated spectrum analysis in the temporal range of -0.1s to 5s was performed by combining the detectors NaI-3,6,7 and BGO-0, {obtaining a spectrum spanning the energy range of} $\sim$ 0.01 - 40 MeV. {As usual}, we {discarded} the 33-37 keV range to avoid the sodium K-edge. The time-integrated spectrum in the given temporal and energy range was extracted utilizing the \sw{GtBurst} software \citep{2023MNRAS.519.3201C, 2024A&A...683A..55C}. To employ the Bayesian analysis method for the temporal and spectral analysis of GBM data, we {used} a Python-based software package, Multi-Mission Maximum Likelihood (\sw{3ML}, \citealt{2015arXiv150708343V}). For the spectral fitting within \sw{3ML}, we utilized a multi-nest sampler with 10,000 iteration steps, as described in {\citet{2018ApJ...864..163V}}. We have used several {empirical models} such as Power Law (PL), \sw{Band}, and \sw{Cutoff PL} (\sw{CPL}) and physical \sw{Synchrotron} and \sw{Blackbody} (\sw{BB}) models and their combinations (\sw{PL+BB}, \sw{Band+BB}, \sw{CPL+BB}), to fit the time-integrated spectrum. \citet{2020NatAs...4..174B} provides the complete details on the physical \sw{synchrotron} model. To compare between the different models, we used the deviance information criteria (DIC).

\begin{figure*}[!t]
\centering
\includegraphics[width=\columnwidth]{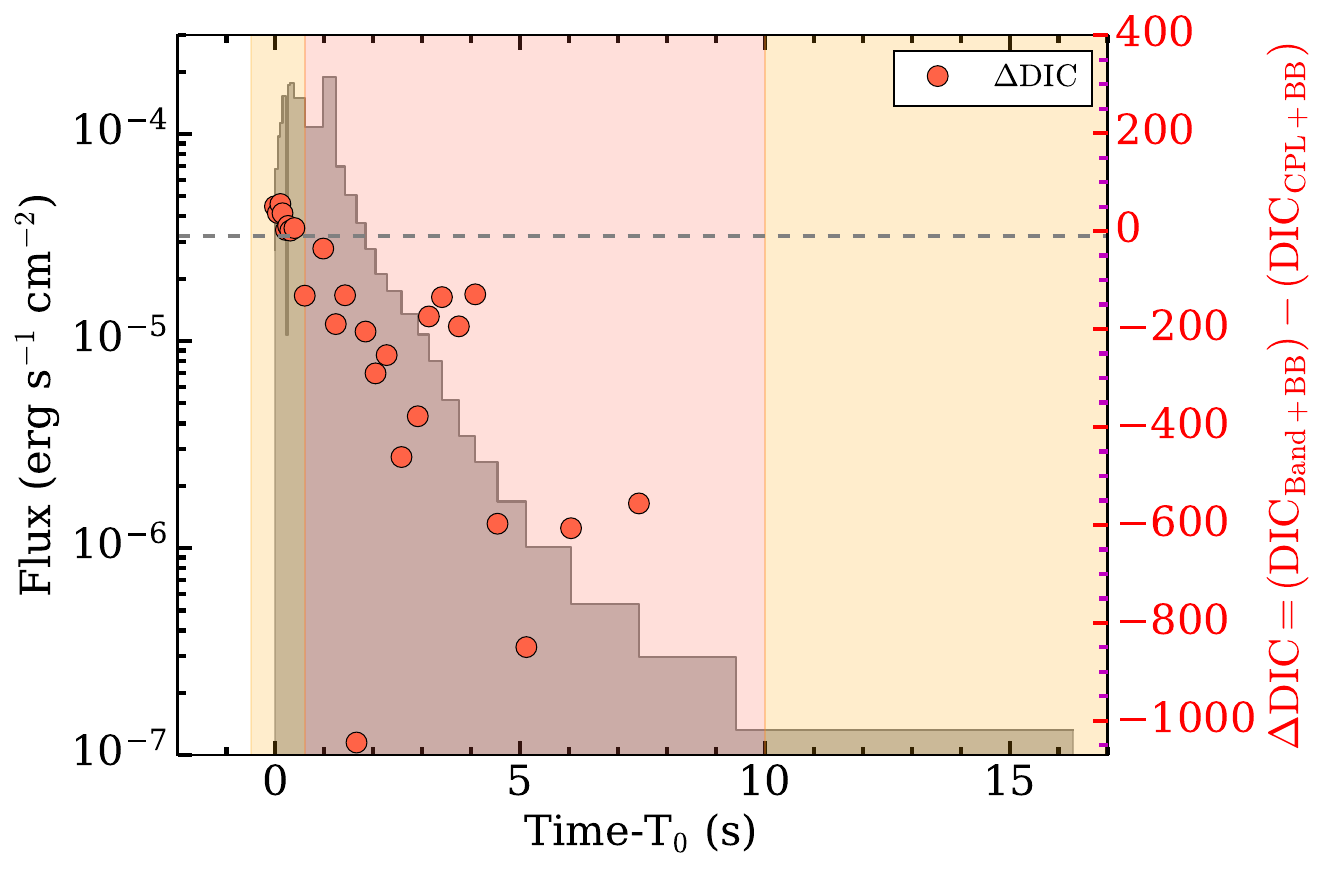}
\includegraphics[width=\columnwidth]{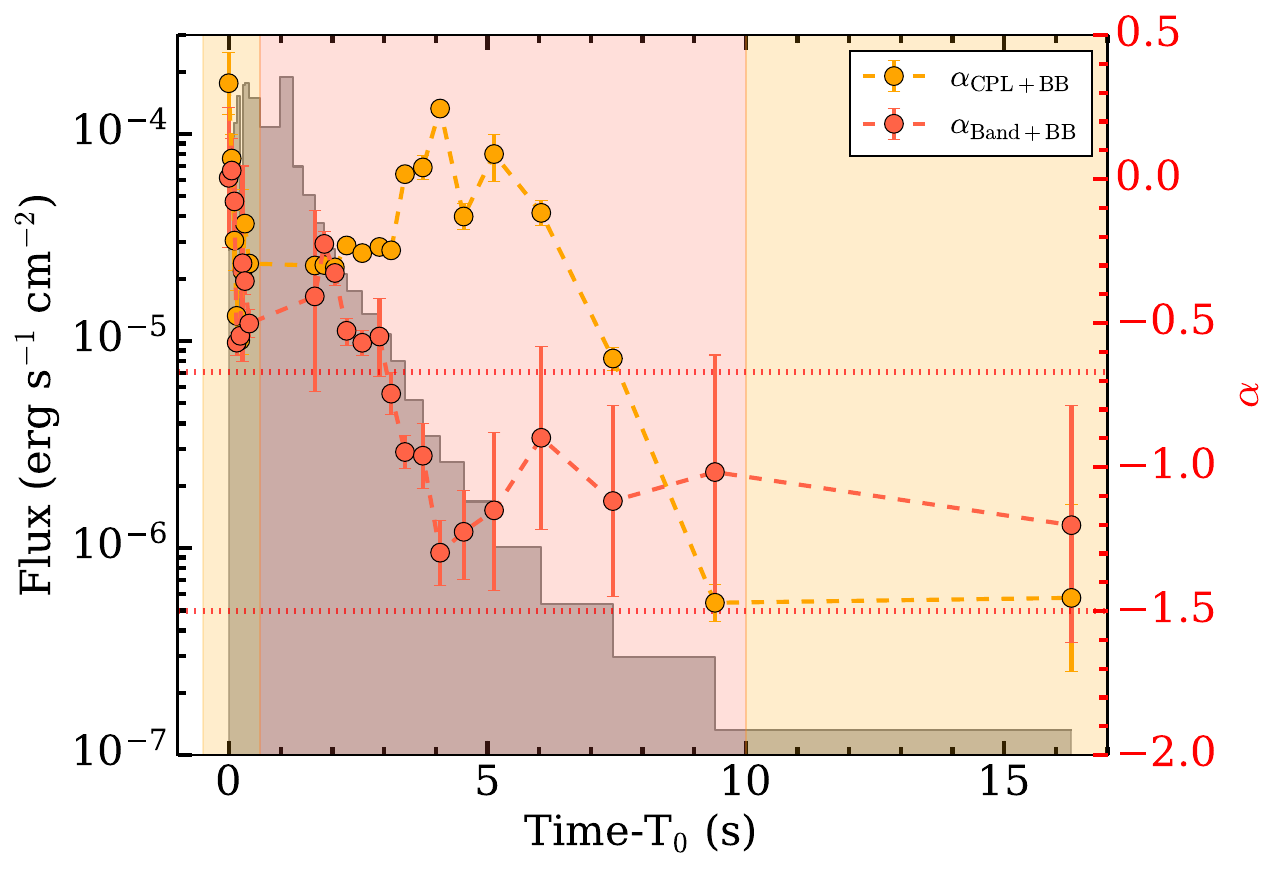}
\includegraphics[width=\columnwidth]{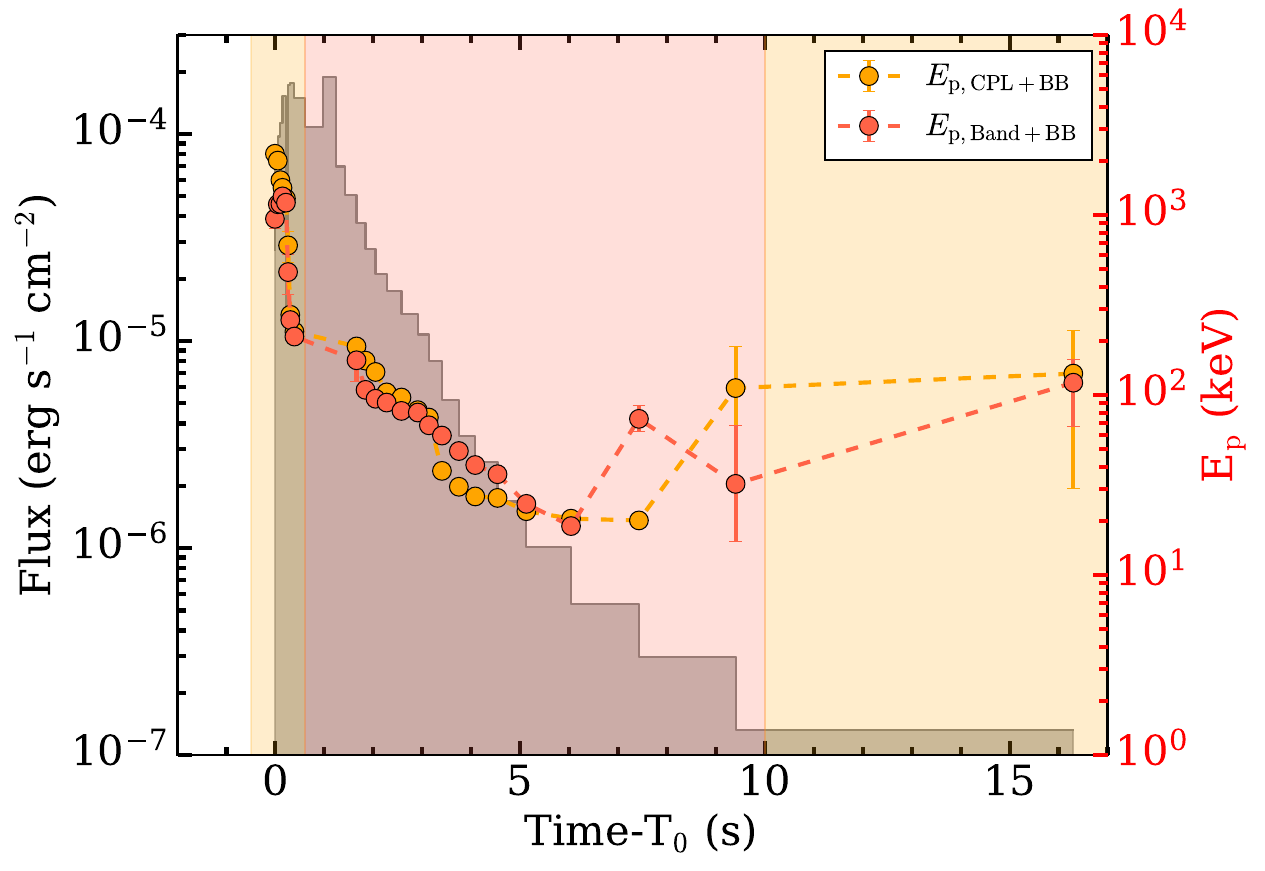}
\includegraphics[width=\columnwidth]{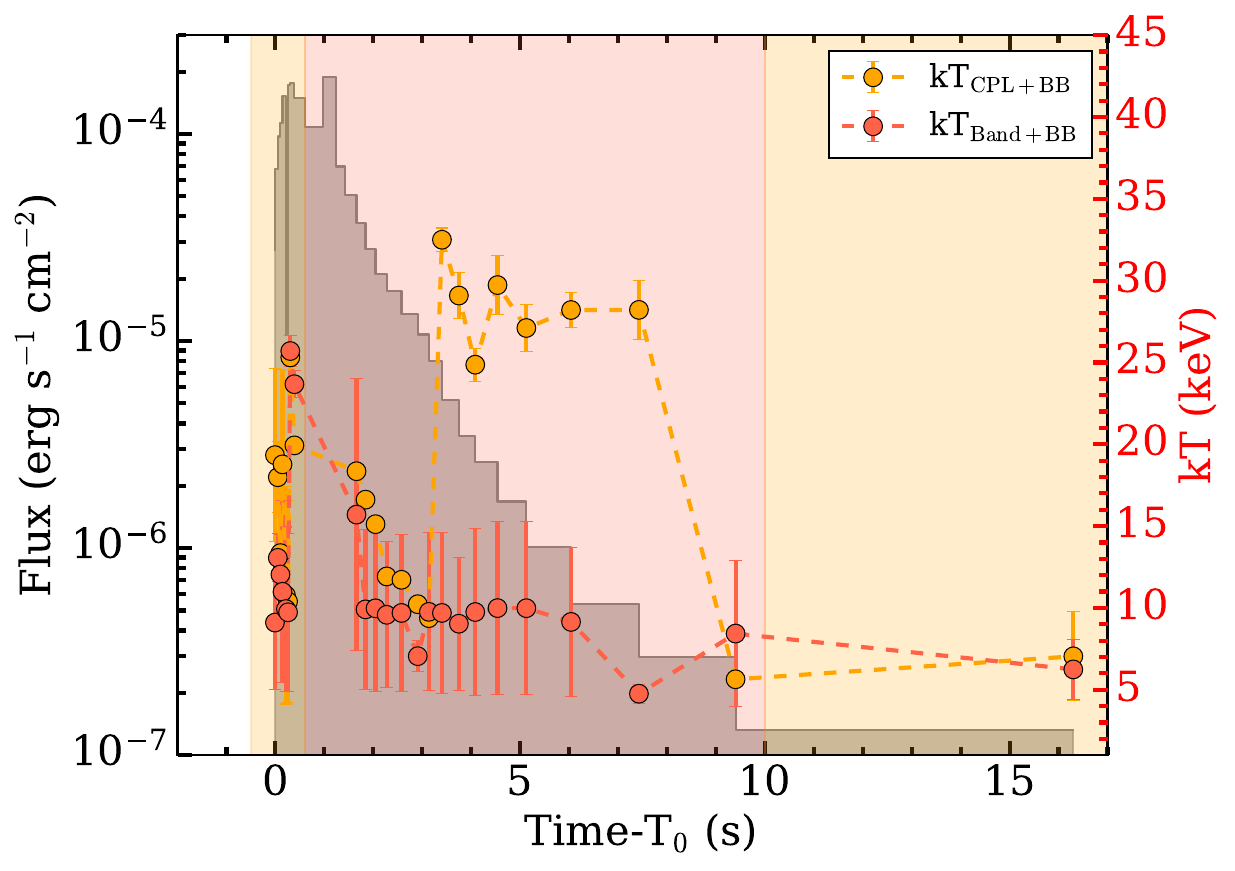}
\includegraphics[width=\columnwidth]{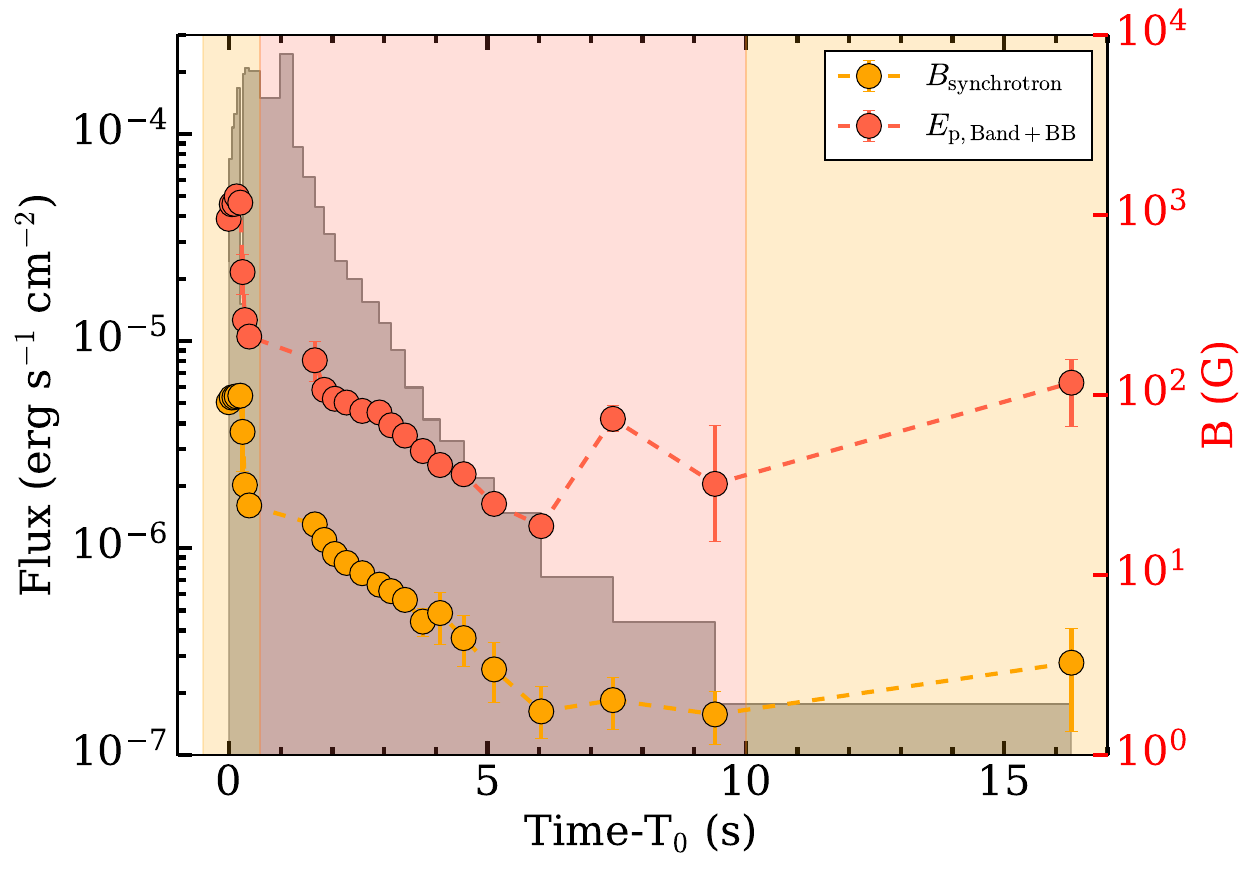}
\includegraphics[width=\columnwidth]{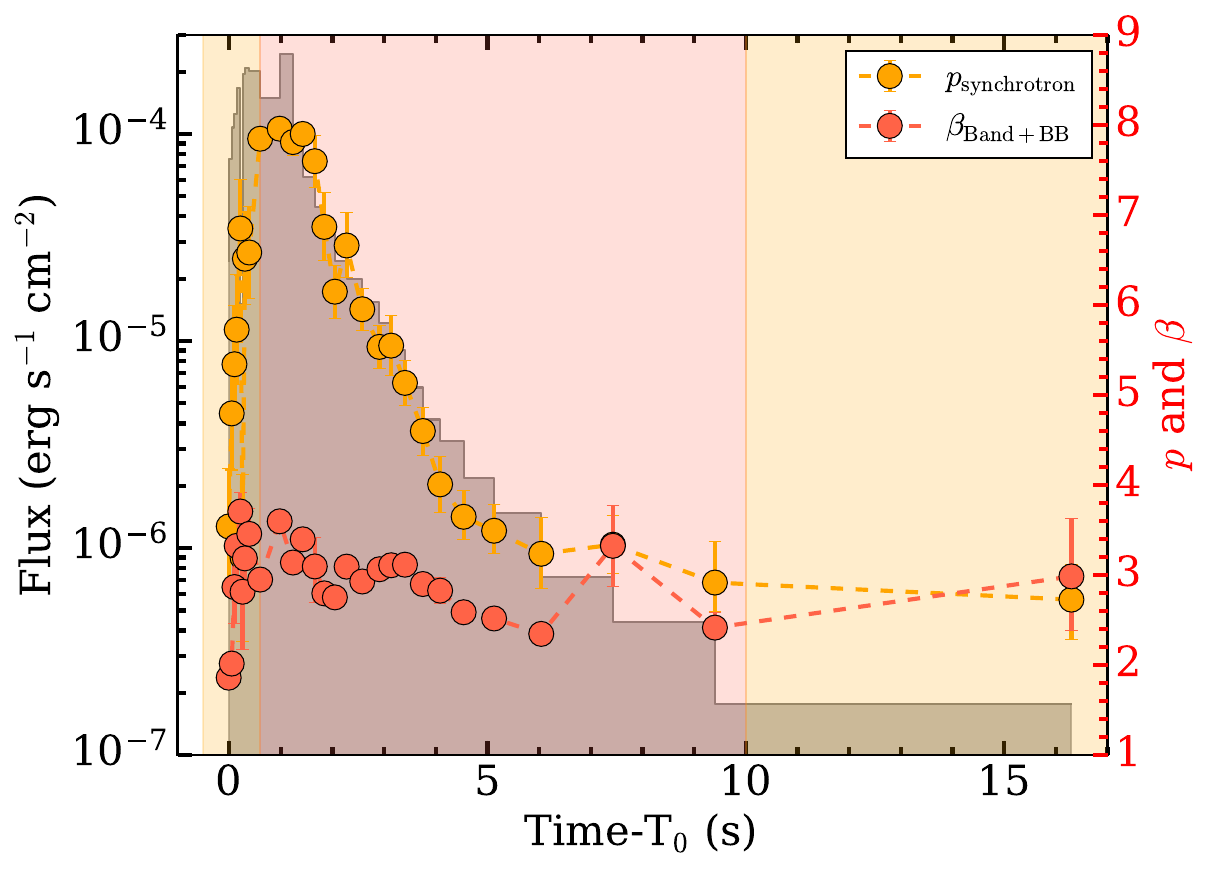}
\caption{Represents the evolution of the prompt emission parameter obtained from the time-resolution spectral analysis of GRB 230812B. In each plot, the left Y-axis represents the flux plotted in the background as a step function, where each step represents the width of the Bayesian block in which the spectral parameters are evaluated, and the X-axis ticks are midpoints of the Bayesian block. The orange-shaded region represents the time range where \sw{CPL+BB} is the best fit, and the red-shaded region is the region where \sw{Band+BB} is the best fit. The difference in the DIC $\Delta$DIC= DIC$_{\rm \sw{Band+BB}}$-DIC$_{\rm \sw{CPL+BB}}$ is shown in the upper-left panel. A dashed line at -10 is plotted to select the best-fit model. The upper-right panel represents the evolution of low energy spectral index $\alpha$ obtained from \sw{CPL+BB} (orange) and \sw{Band+BB} (red). Two dotted lines at -3/2 and -2/3 represent the limits of \sw{synchrotron} emission from external shock \citep{2000ApJS..126...19P}. Similarly, the middle-left and middle-right panels represent the evolution of E$_{\rm p}$ and kT, respectively. The lower two panels represent the magnetic field strength B and the energy distribution index $p$ of the shock-accelerated electrons.}
\label{fig:param_evolution}
\end{figure*}

Further, we performed the time-resolved spectral analysis of \fermi-GBM data to study the {parameter's} evolution during the prompt emission. For the time-resolved spectral analysis, the GBM LC from the brightest detector (NaI-6) was rebinned using the methodology known as Bayesian block binning by setting the false alarm probability at 0.01 \citep{2013ApJ...764..167S}. The obtained time slices were then applied to all other detectors. The advantages of Bayesian block binning methods are given in \citet{2014MNRAS.445.2589B}. From this method, we obtained 29 spectra, out of which only 24 are statistically significant with S $>$ 20 \citep{2018ApJ...864..163V}. After extracting the spectrum using \sw{GtBurst}, we fit each spectrum with several {empirical} and physical models as discussed above for time-integrated analysis. The results of the spectral analysis for the prompt emission are presented in Figure \ref{fig:param_evolution} and elaborated upon in section \ref{sec:results}.

\subsection{Afterglow observation and analysis}
Initially, the burst was beyond the observational range of the Burst Alert Telescope onboard {\em Neil Gehrel Swift Observatory} (\swift-BAT, \citealt{2005SSRv..120..143B}). Following a sequence of tiled observations, the \swift X-ray Telescope (XRT) successfully localized the burst at \tzero+25 ks \citep{2023GCN.34388....1E, 2023GCN.34393....1K, 2023GCN.34394....1P}. All \swift-XRT observations were taken in photon count (PC) mode. The preliminary XRT LC and spectrum {were modeled} by a simple PL and absorbed PL with indices $\alpha_{\rm x}$ = 1.80 $\pm$ 0.4 and $\Gamma_{\rm x}$ ($\beta_{\rm x}$+1) = 1.82 $\pm$ 0.15 \citep{2023GCN.34400....1B}.

To delve into the afterglow characteristic of GRB 230812B, the X-ray temporal and spectral data {were} obtained from the online repository \footnote{\url{https://www.swift.ac.uk/xrt_curves/00021589/}} of UK \swift Science Data Centre {\citep{2007A&A...469..379E, 2009MNRAS.397.1177E}}. For {the optical} data analysis, we obtained the data reported {from} GCN circulars \footnote{\url{https://gcn.nasa.gov/circulars}}. The redshift ($z$=0.36) of the burst was measured from the spectroscopic {observations using the} 10.4m Gran Telescopio d{e Canarias} \citep{2023GCN.34409....1D}.

\begin{figure*}[!t]
\centering
\includegraphics[width=\textwidth]{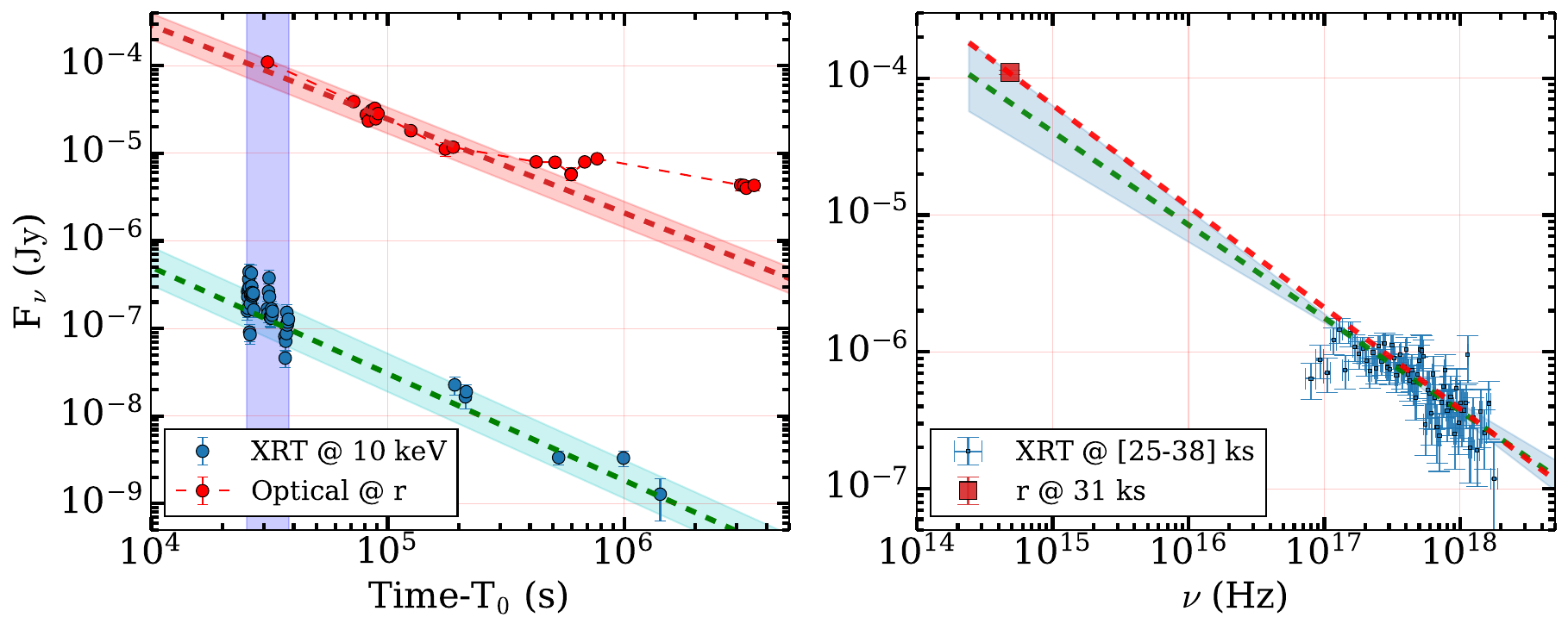}
\caption{Left panel: XRT (green) and r-band optical LCs (red) are plotted. The colored dotted line and the shaded area around them represent the PL fit and associated error bar for both LCs. The vertically shaded region corresponds to the time interval used to construct an optical r-band to X-ray SED. Right pane: optical to X-ray SED. Optical r-band data is plotted in a red square, and X-ray observation is shown in blue. The blue dotted line represents the PL fitted to X-ray spectra at 0.3-10 keV. The red dotted line represents the PL model fitted to the combined r-band to X-ray SED. Green squares and the dotted line represent the LAT spectra and corresponding power law fit, respectively.}
\label{fig:sed}
\end{figure*}

In the case of GRB 230812B, the flux density X-ray LC at 10 keV showed no flare or plateau throughout {the XRT observations} from \tzero+25 ks to \tzero+1400 ks. We modeled the X-ray LC with a simple PL using the MCMC technique. We obtained a PL {index of} $\alpha_{\rm x}$ = -1.22$_{-0.05}^{+0.05}$. {The X-ray }flux density LC and the PL model fitted to it are shown in the left panel of Figure \ref{fig:sed}.

Similarly, the optical LC also seems to decay without breaks. Following a similar methodology, we fitted the optical LC using a simple PL and obtained the decay index of $\alpha_{\rm o}$ = 1.08$_{-0.03}^{0.03}$. As shown in the left {panel of} Figure \ref{fig:sed}, the optical LC starts deviating from the PL decay at $\sim$ 4 days after \tzero. This is {because the} underlying supernova emission began to emerge about four days after the burst trigger {\citep{2024ApJ...960L..18S, 2023arXiv231014310H}}.

We have retrieved the X-ray spectra in the {time range} of 25 ks to 38 ks and the {energy range of} 0.3 - 10 keV. {For the} XRT spectrum fitting, we utilized an absorbed PL model with a multiplicative Galactic absorption component {(\textit{phabs})} and the host absorption component {(\textit{zphabs})} in the X-ray spectral fitting package (\sw{XSPEC}; \citealt{1996ASPC..101...17A, 2022JApA...43...11G}). {The galactic} hydrogen column density {(NH$_{\rm Gal}$)} was fixed at 2.02 $\times$ 10$^{20}$ cm$^{-2}$ during the spectral fitting. There is no flare or {any other} variation present in the XRT LC. We first calculated the host hydrogen column density (NH$_{\rm z}$ = 7.20 $\times$ 10$^{20}$ cm$^{-2}$) along the line {of sight} by fitting the {absorbed PL} to the average spectrum. Then, by fixing both NH$_{\rm Gal}$ and NH$_{\rm z}$, we again fit the XRT spectrum with the same absorbed PL model and obtained a spectral {index of} $\beta_{x}$ = 0.68$_{-0.08}^{+0.08}$. 

After that, we created an optical r-band to X-ray spectral energy distribution (SED) {in the time interval} 25 ks - 38 ks. We fit the optical-X-ray SED with the PL and obtained {a} PL index $\beta_{\rm ox}$ = 0.74$_{0.01}^{0.01}$. {The optical} r-band to X-ray SED and corresponding PL fit is presented in the right panel of Figure \ref{fig:sed}.

\subsection{\fermi-LAT observation and analysis}
At the moment of {the} GBM detection, the LAT boresight angle was 29$^{\circ}$. \citet{2023GCN.34392....1S} reported that \fermi-LAT detected the GeV photons from the GRB 230812B. Preliminary LAT spectral analysis above 100 MeV revealed the PL distribution of energy with the photon index $\Gamma_{\rm LAT}$ = 2.16 $\pm$ 0.14. \fermi-LAT recorded a photon with a maximum energy of 72 GeV at \tzero+32.2s \citep{2023GCN.34392....1S}.

We employed the \sw{GtBurst} software to download \fermi-LAT data spanning \tzero to \tzero+10 ks. Subsequently, we conducted an unbinned likelihood analysis on the time-integrated LAT data, covering the energy range of 0.1-100 GeV. In our analysis of LAT data, we selected a 10$\times$10 {deg$^{2}$} region centered around the GRB 230812B. The instrumental response function P8R3\_SOURCE, a maximum zenith cut of 100 degrees, and a skybinning of 0.2 degrees were applied. Finally, we utilized the \sw{gtsrcprob} tool to calculate the probability of observed photons associated with GRB 230812B.

\begin{figure}[!t]
\hspace{-0.5cm}
\includegraphics[width=\columnwidth]{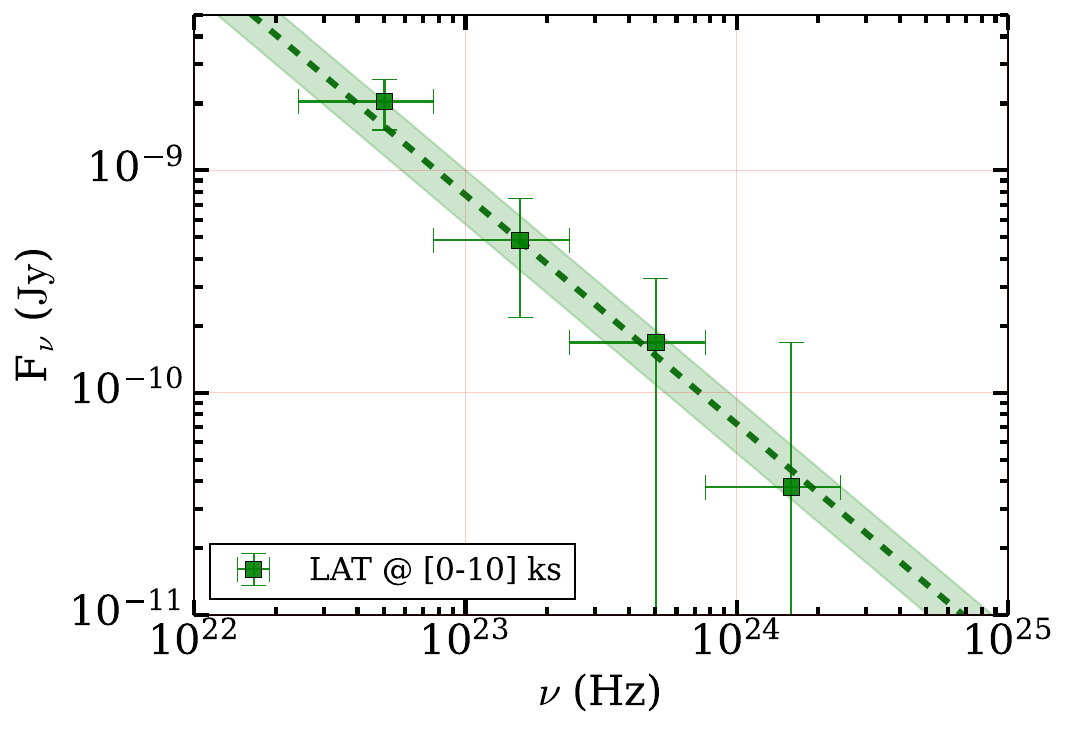}
\caption{Represents the fitting results of time average LAT spectra in the energy range 0.1-100 GeV}
\label{fig:latspec}
\end{figure}

The time-integrated spectral analysis of LAT data yielded a photon index $\Gamma_{\rm LAT}$ = -1.99 $\pm$ 0.13, with an average flux = 3.79 $\pm$ 0.68 $\times$ 10$^{-9}$ erg s$^{-1}$ cm$^{-2}$ in the 0.1-10 GeV energy range. Additionally, we extracted the spectral file for \sw{XSPEC}. We again fit the extracted spectra in \sw{XSPEC} and obtained a photon index of $\Gamma_{\rm LAT}$ = 2.03$_{-0.23}^{+0.27}$ and a flux of 3.52 $\pm$ 0.98 $\times$ 10$^{-9}$ erg s$^{-1}$ cm$^{-2}$, consistent within the error. We saved the QDP file from the \sw{XSPEC} and plotted the LAT spectra in Figure \ref{fig:latspec}.

Following this, we performed the time-resolved analysis of LAT data. Initially, we binned the LAT data in 5 s bins, and subsequently, the bin width was increased, as illustrated in Figure \ref{fig:latlc}. Up to the initial 100 s, the instrument response function \sw{P8R3\_TRANSIENT020} was utilized, followed by a switch to the instrument response function \sw{P8R3\_SOURCE}. The outcomes of the time-resolved spectral analysis are presented in Figure \ref{fig:latlc}. The flux obtained in the range 0.1-10 GeV decreased with time following a PL decay, with the decay index $\alpha_{\rm LAT}$ = 1.04 $\pm$ 0.08.

\section{Results} \label{sec:results}
\subsection{Prompt emission}
The prompt emission multichannel LC of GRB 230812B is presented in the upper and middle panels of Figure \ref{fig:mclc}. The evolution of hardness ratio (HR, {i.e.} no. of photons in 50-300\,keV / no. of photons in 8-50\,keV) is presented in the lower panel of Figure \ref{fig:mclc}. The prompt emission LC is the fast rise and exponential decay (FRED) type with a single broad pulse. {The HR} remains $>$ 1 for almost \tzero to \tzero+ 2\,s and seems to follow the intensity of the burst, {indicating that} GRB 230812B is a hard burst.

{The} time-integrated spectrum of GRB 230812B is best modeled by \sw{Band+BB} with the minimum DIC value of 9531 over \sw{synchrotron}, Band, \sw{CPL}, and \sw{CPL+BB} with DIC values of 9570, 10356, 24725, and 13234, respectively. {The} best-fit parameter obtained from the best-fit model are $\alpha$ = -1.02$_{-0.01}^{+0.01}$, \Ep = 411.35$_{-8.18}^{+8.21}$\,keV, $\beta$ = -2.75$_{-0.04}^{+0.04}$ and kT = 24.05$_{-0.32}^{+0.32}$\,keV.

The prompt emission time-resolved spectral fitting results are presented in Figure \ref{fig:param_evolution}. The upper left panel of Figure \ref{fig:param_evolution} illustrates the difference between the DIC values of two best-fit models, \sw{Band+BB} and \sw{CPL+BB}, i.e., $\Delta$DIC = DIC$_{\rm \sw{Band+BB}}$-DIC$_{\rm \sw{CPL+BB}}$. $\Delta$DIC $<$ -10 {means} {that the} \sw{Band+BB} is best fitting; otherwise, \sw{CPL+BB} is best fitting. {The} time-resolved spectral analysis revealed that during the rising phase of the {LC's} prompt emission, {the} time-resolved spectra are best fitted by the \sw{CPL+BB} model. Conversely, during the decay phase of the LC, all the spectra are best fitted by the \sw{Band+BB} model based on the lower DIC values. In Figure \ref{fig:param_evolution}, the orange shade region represents the interval during which \sw{CPL+BB} is best fitting, and the red shaded region represents the interval during which \sw{Band+BB} best fits the time-resolved spectra. The total isotropic $\gamma$-ray energy release $E_{\gamma, iso}$ = 8.74 $\pm$ 1.61 $\times$ 10$^{52}$ erg along {with the} \Ep value is found {to be} consistent with the Amati correlation for LGRBs {\citep{2006MNRAS.372..233A}}.

\subsection{The prompt emission {parameter's} evolution}
The evolution of the low energy spectral index $\alpha$ obtained from the best-fit model is presented in the upper right panel of Figure \ref{fig:param_evolution}. Throughout the prompt emission, $\alpha$ seems to follow the intensity. The $\alpha$ obtained from the \sw{CPL+BB} {in the} 2-10\,s {time intervals} deviates from {the flux} tracking. This might be due to the \sw{CPL+BB} model not fitting well in the given {time interval}. Similar to the $\alpha$, the peak energy \Ep of the spectrum is also found to be following the intensity. Hence, {the} GRB 230812B is a double-tracking burst. The time-integrated and time-resolved spectral analysis results showed that the single empirical \sw{Band} or \sw{CPL} {do not fit well the spectra}, but an addition of a thermal component is required. However, {the} evolution of the thermal energy component kT is somewhat complicated. It closely follows the flux up to \tzero+3 s. After that, it shows almost a constant value of $\sim$ {7\,keV}, but with larger error bars. {Nevertheless}, the physical \sw{Synchrotron} model did not fit the spectrum well. However, it is still reasonable to check for the evolution of {the magnetic} field strength (B) obtained from {fitting with the} \sw{Synchrotron} model. The evolution of B closely matches with the evolution of \Ep, as shown in the lower panel of Figure \ref{fig:param_evolution}. {The} electron energy distribution in the \sw{synchrotron} model follows a PL with a PL index $p$. {The parameter $p$ is obtained} from fitting the \sw{synchrotron} model {and} is also showing a flux tracking evolution as shown in the lower right panel of Figure \ref{fig:param_evolution}.

\subsection{Afterglow emission and the closure relations}
The observed X-ray and optical r-band {LCs} of GRB 230812B do not show any flare, plateau, or bump (see Figure \ref{fig:sed}). Both X-ray and r-band LCs were fitted with a simple PL {obtaining} temporal decay {indices of} $\alpha_{\rm x}$ = -1.22$_{-0.05}^{+0.05}$ and $\alpha_{\rm o}$ = 1.08$_{-0.03}^{0.03}$. Both X-ray and optical decay indices are consistent with each other within 3$\sigma$. In addition, {the} spectral indices $\beta_{\rm x}$ = 0.68$_{-0.08}^{+0.08}$ and $\beta_{\rm ox}$ = 0.74$_{0.01}^{0.01}$ obtained from the fitting of XRT spectra in 0.3-10 keV and combined optical r-band to X-ray SED are also consistent within 1$\sigma$. This indicates that both X-ray and optical emissions come from the same spectral regime. We {checked} the closure relation followed by the GRB external shock \citep{1998ApJ...497L..17S}. We {found that the} optical and X-ray spectral and temporal indices satisfy the relation $\beta$ = $\frac{p - 1}{2}$ and $\alpha$ = $\frac{3(p-1)}{4}$ in the spectral regime $\nu_{\rm m}$ $<$ $\nu_{\rm o}$ $<$ $\nu_{\rm x}$ $<$ $\nu_{\rm c}$ in the ISM-like surrounding medium.

\fermi-LAT data was not included in the Spectral Energy Distribution (SED) due to the absence of optical and X-ray observations during the \fermi-LAT detection. From \fermi-LAT temporal and spectral analysis we obtained decay indices {of} $\alpha_{\rm LAT}$ = 1.04 $\pm$ 0.08 and $\beta_{\rm LAT}$ ($\Gamma_{\rm LAT}$-1) = 1.03$_{-0.23}^{+0.27}$. {The} observed $\beta_{\rm LAT}$ is slightly steeper than $\beta_{\rm o}$ and $\beta_{\rm x}$. The difference $\Delta{\beta}$ = $\beta_{\rm LAT}$-$\beta_{\rm x}$ is 0.4 $\pm$ 0.3, which is consistent with the assumption that $\nu_{c}$ might {lie} between the X-ray and LAT observation in the ISM medium.

\begin{figure}[!t]
\hspace{-0.5cm}
\includegraphics[width=1.1\columnwidth]{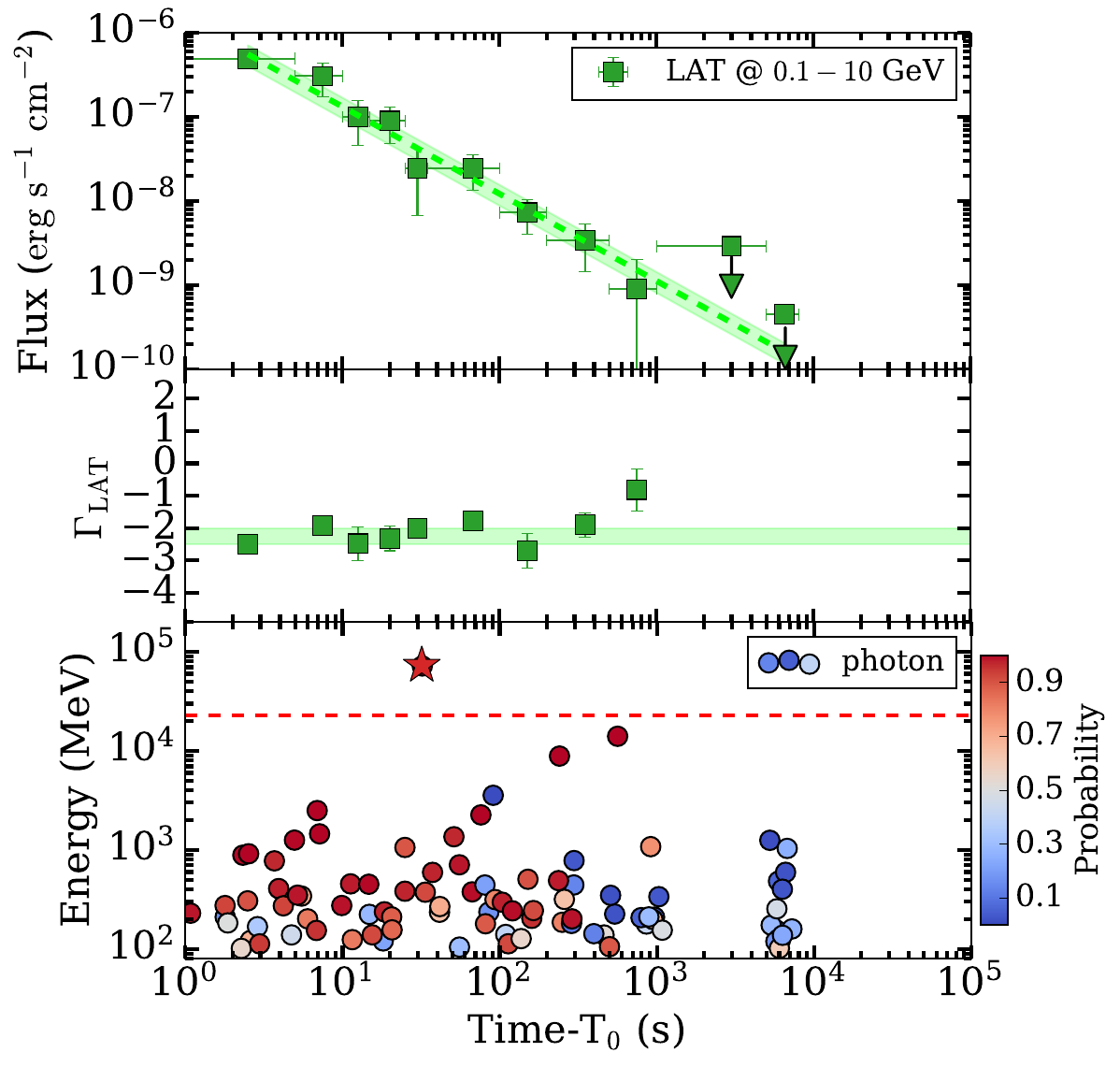}
\caption{Top panel: represents the \fermi-LAT LC of GRB 230812B in the energy range of 0.1-10 GeV. Squares denote the observed flux, and the dotted line represents the PL fitted to the LC. The shaded area represents the 1$\sigma$ error associated with the fit. The middle panel represents the evolution of the photon index derived from the time-resolved fitting of LAT spectra. The bottom panel displays the temporal distribution of LAT-detected photons, with the color map indicating the probability of the photons being associated with GRB 230812B. The red star at 32 s represents the maximum energy photon of 72 GeV from \fermi-LAT. The red dotted line represents the maximum energy allowed by synchrotron emission for GRB 230812B.}
\label{fig:latlc}
\end{figure}

\subsection{Discussion}
\subsection{Afterglow LC comparison}

\begin{figure*}[!t]
\centering
\includegraphics[width=\columnwidth]{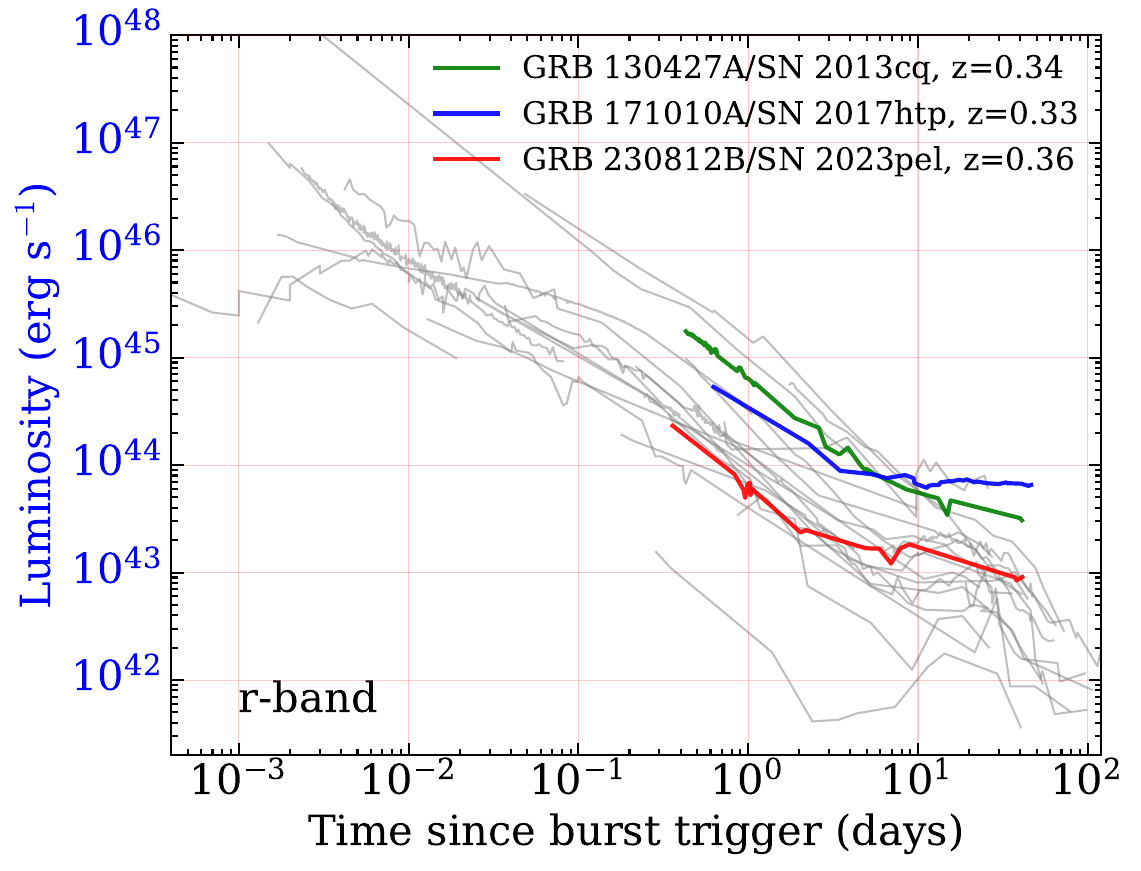}
\includegraphics[width=\columnwidth]{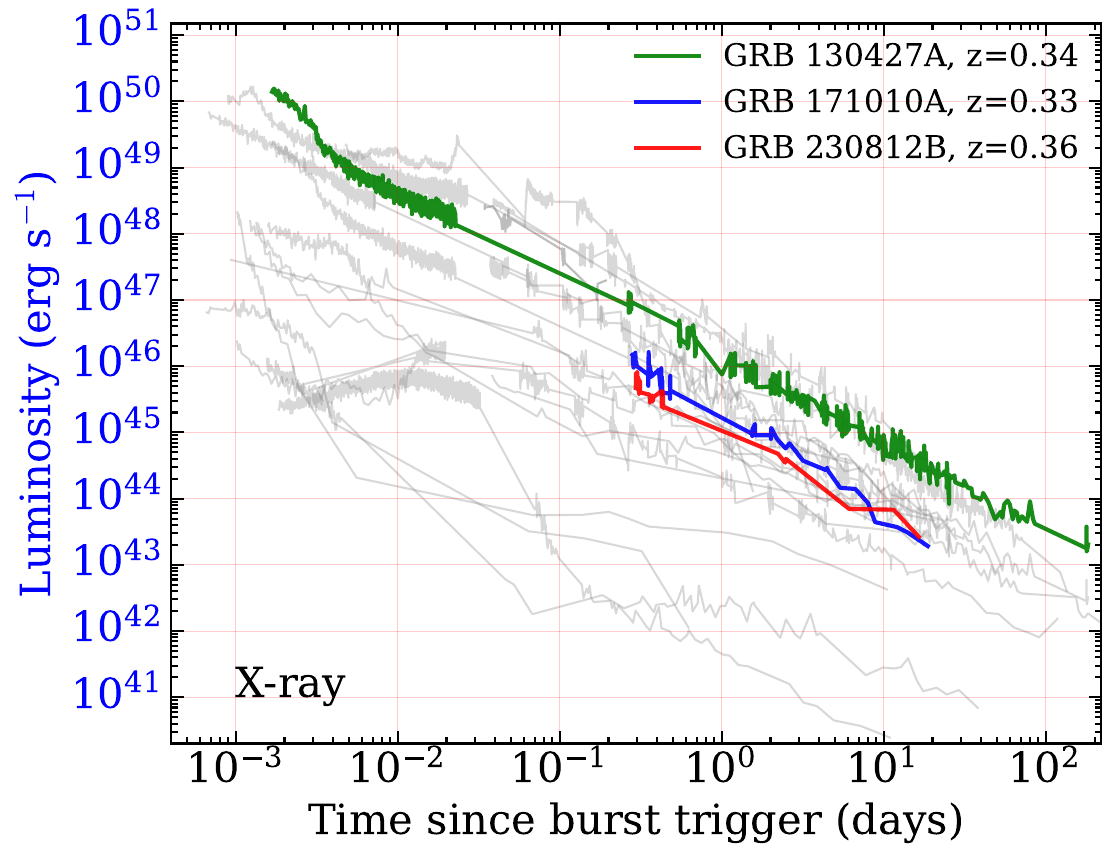}
\caption{A comparison of optical r-band (left) and \swift-XRT (right) luminosity LCs of GRB 130427A, GRB 171010A, and GRB 230812B (red) with other SN-detected bursts (grey) in the background, see \citet{2023ApJ...942...34R} for more details.}
\label{fig:r_x_cmp}
\end{figure*}

We compared the X-ray and optical afterglow LCs of GRB 230812B with a set of 15 SN-detected GRB from \citet{2022NewA...9701889K} in Figure \ref{fig:r_x_cmp}. We found the \swift-XRT and optical r-band afterglow LC of GRB 230812B is consistent with other GRBs in the background and does not possess an exceptional bright afterglow or SN emissions. Even the optical and X-ray LC of GRB 230812B is relatively fainter than the bright {bursts of} GRB 130427A \citep{2013ApJ...776...98X} and GRB 171010A \citep{2022NewA...9701889K} at similar redshifts as shown in Figure \ref{fig:r_x_cmp}. Our analysis concluded that the afterglow of GRB 230812B and its associated SN 2023pel \citep{2023arXiv231014310H, 2024ApJ...960L..18S} is an intermediate bright GRB/SN compared to {other} GRBs.

\subsection{Possible origin of GeV emission from \fermi-LAT} 
The observed power-law decay of the LAT light curve and spectra {indicate} that the LAT emission originates from an external shock. However, unlike optical and X-ray observations, the emission mechanism of LAT-detected photons still needs to be completely {clarified}. According to {\citet{2019ApJ...879L..26F}}, the maximum energy of {photons} from \sw{synchrotron} external shock model can be 10 GeV $\times$ ($\Gamma_{0}$/100) $\times$ (1+z)$^{-1}$. Where $\Gamma_{0}$ is {the} bulk Lorentz factor, and z is redshift. Considering the bulk Lorentz factor $\Gamma_{0}$ = 315, calculated {it} from the observed E$_{\rm \gamma, iso}$ using the relation given in {\citet{2010ApJ...725.2209L}}. Using the above relations, we calculated that, in the case of GRB 230812B, the \sw{Synchrotron} emission mechanism could only generate a maximum energy photon of 23 GeV, as shown with a red dotted line in the lower panel of the Figure \ref{fig:latlc}. From the lower panel of Figure \ref{fig:latlc}, it is clear that all photons lie below this level except for 72 GeV {photons}. Thus, the origin of the 72 GeV {photons} cannot be from \sw{synchrotron} emission. This has been observed earlier for {the} VHE bursts (GRB 180720B, 190114C, 190829B, GRB 201015A, GRB 201216C: \citealt{2019ApJ...879L..26F, 2019Natur.575..459M, 2021ApJ...918...12F, 2021Sci...372.1081H, 2023ApJ...942...34R}), where the observed VHE photons are explained via \sw{synchrotron self Compton (SSC)} or external \sw{Inverse Compton (IC)}. It is possible that the non-thermal photons from {the prompt} emission {were} Lorentz boosted by accelerated electrons in the external shock {resulting} in the observed LAT emission. This suggests that the \sw{IC} and \sw{SSC} are the potential emission mechanisms that could be responsible for the VHE 72 GeV photons observed from GRB 230812B.

\subsection{GRB 230812B compared with other S- and LGRBs} 
\begin{figure*}[t]
\centering
\includegraphics[width=0.95\columnwidth]{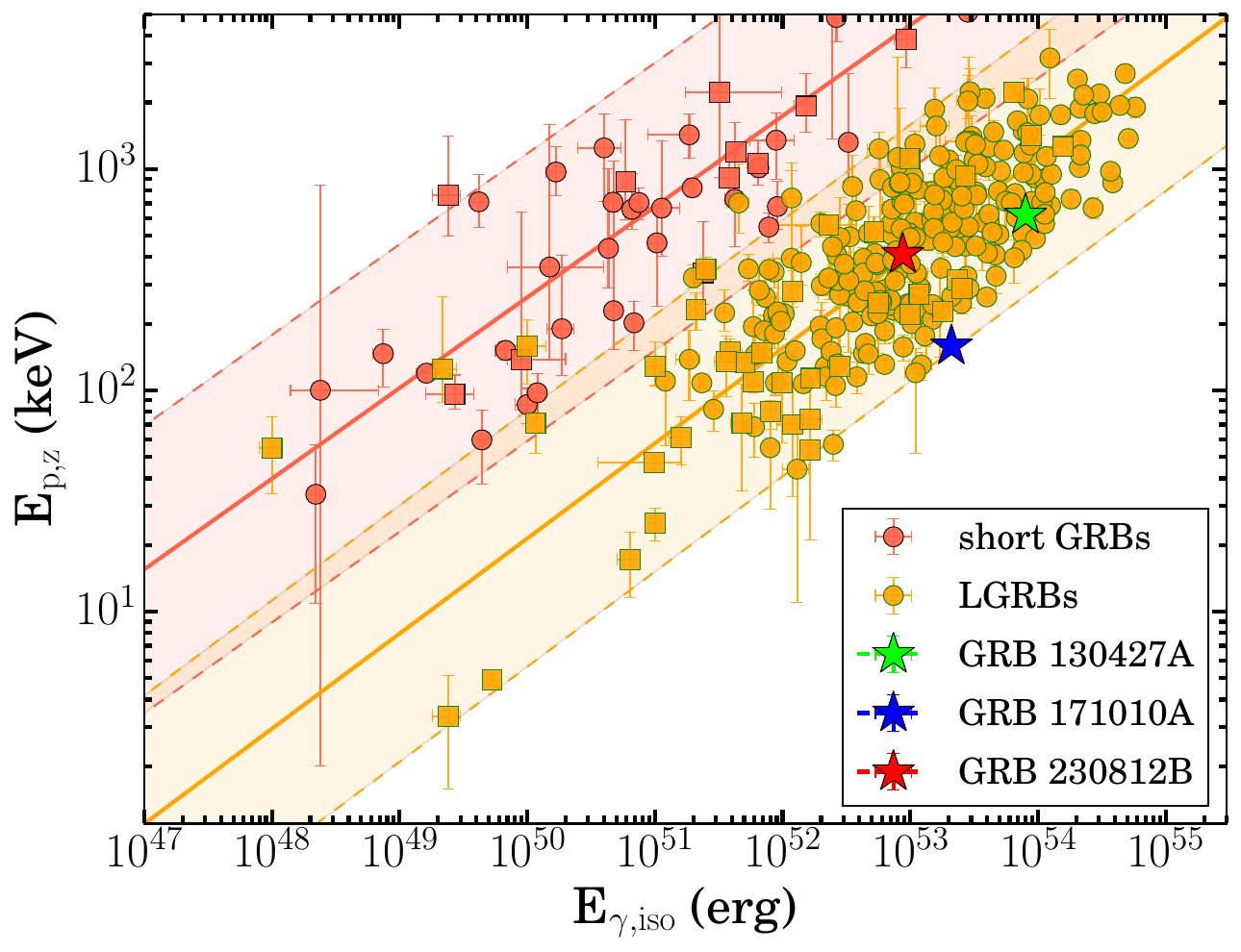}
\includegraphics[width=1.06\columnwidth]{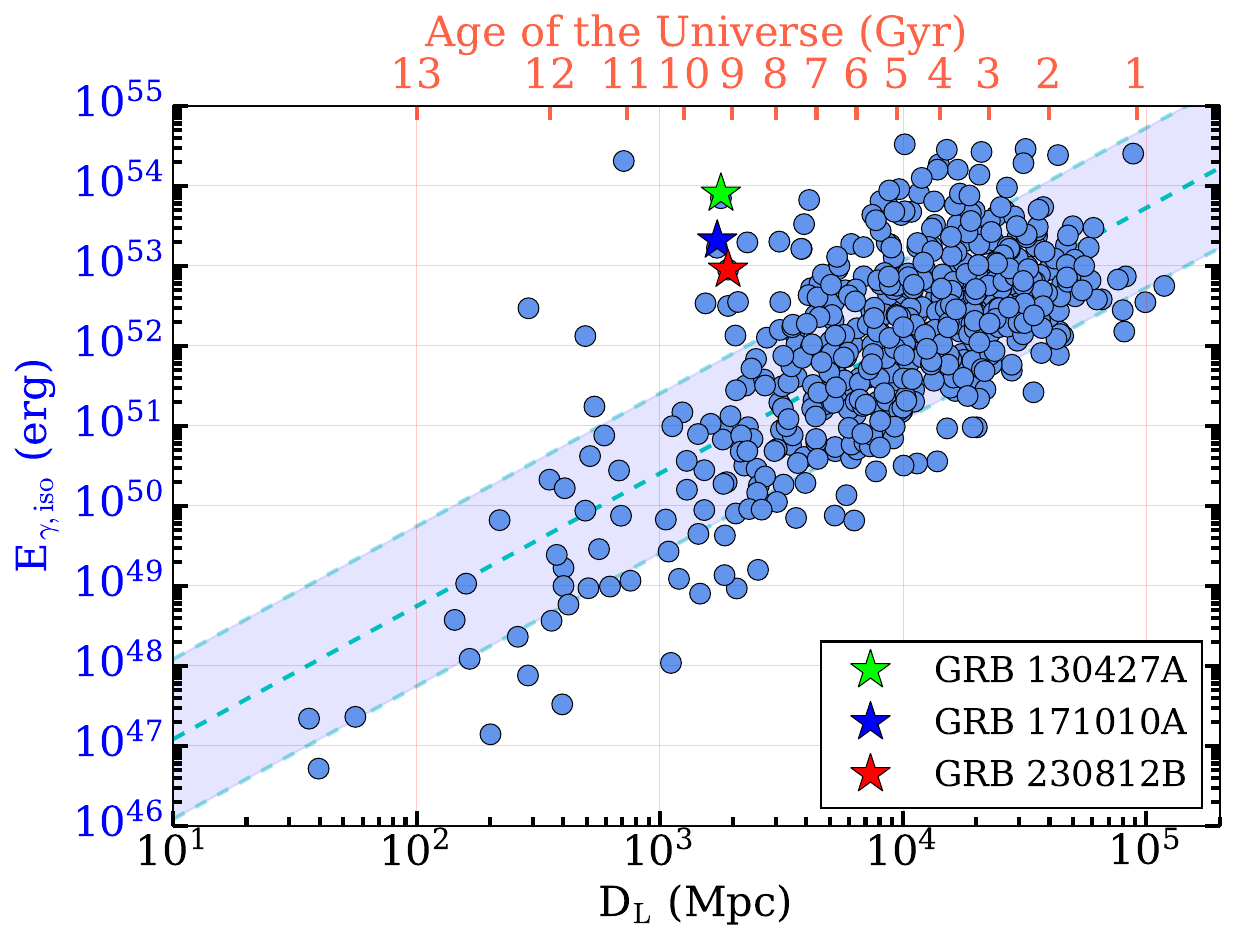}
\includegraphics[width=0.95\columnwidth]{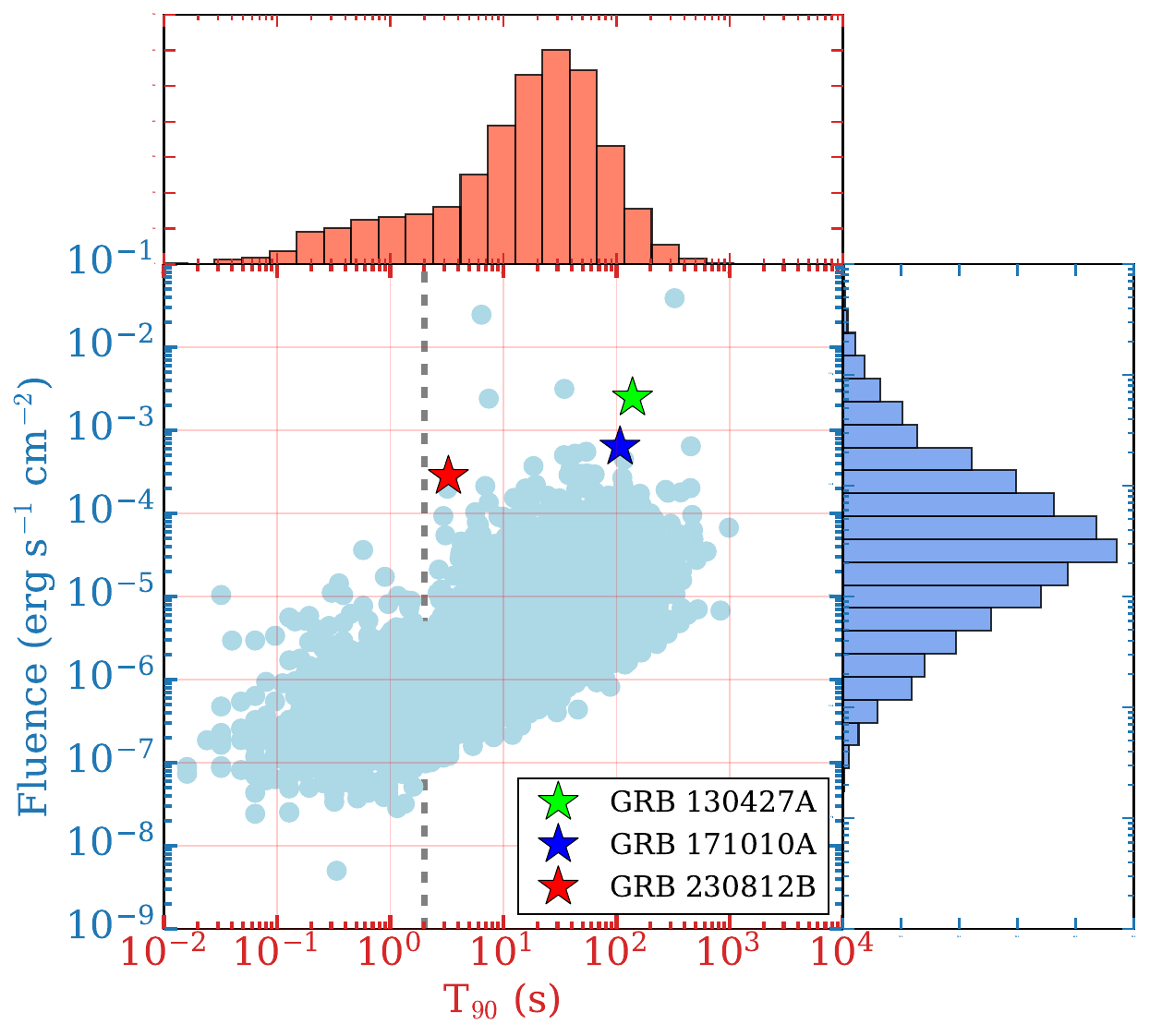}
\includegraphics[width=1.15\columnwidth]{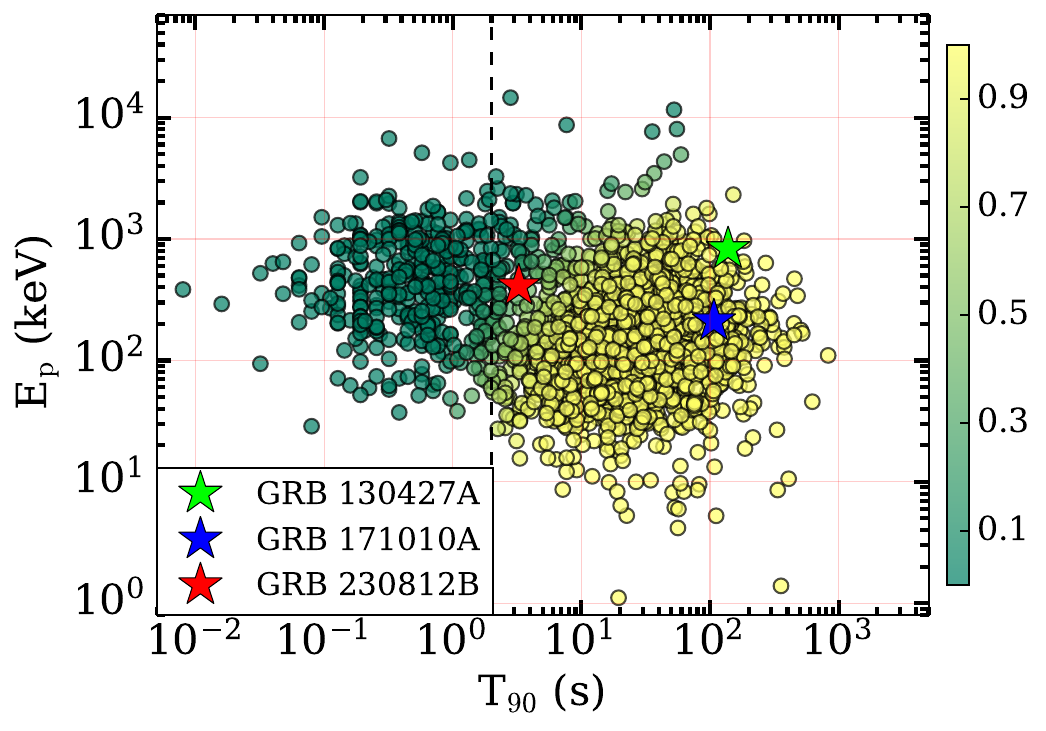}
\caption{Prompt emission characteristic of GRB 230812B compared with SN-detected GRB 130427A and GRB 171010A at similar redshift. Upper left panel: represents GRB 230812B in Amati correlation space \citep{2006MNRAS.372..233A}. Upper right panel: represents the E$_{\rm \gamma,iso}$ distribution of GRBs plotted against the luminosity distance in Mpc and the universe's age in Gyr. Lower left panel: represents the fluence distribution of GRBs along with the T$_{90}$ duration on the observer frame. Lower right panel: \Ep in the observed frames plotted with the \tninty duration. A vertical dashed line at 2s separates between SGRBs and LGRBs.}
\label{fig:DL}
\end{figure*}

The {characteristics} of GRB 230812B compared to the other burst from the IceCube\footnote{\url{https://user-web.icecube.wisc.edu/~grbweb_public/Summary_table.html}} catalog are shown in the Figure \ref{fig:DL}. In the upper right panel of Figure \ref{fig:DL}, we compared the E$_{\rm \gamma, iso}$ of redshift detected bursts. The observed E$_{\rm \gamma, iso}$ of the bursts is found to correlate with the luminosity distance (D$_{L}$). We fitted a PL and found the relation:
$ \rm log_{10}(E_{\rm \gamma,iso}) = (45.42\pm0.16) + (1.66\pm0.04) \times log_{10}(D_{L})$.

GRB 230812B, with the observed distance of {1.9 Gpc} corresponding to the cosmological age of 9.75 Gyr, is inconsistent with the above relation. Along with GRB 130427A and GRB 171010A, GRB 230812B lies outside of the 5$\sigma$ range of relations, as shown in Figure \ref{fig:DL}. This is due to its exceptionally high prompt emission brightness at z=0.36. Most of the other GRBs at this redshift are fainter than GRB 230812B, except for GRB 130427A and GRB 171010A. In the lower-left panel of Figure \ref{fig:DL}, we compared the observed fluence of GRB 230812B (red star), GRB 130427A and GRB 171010A along with other bursts from the IceCube catalog. GRB 230812B is the brightest burst with \tninty $<$ 5 s, and only slightly fainter than GRB 130427A and GRB 171010A at the similar redshift. This huge flux detected {in such} sort time scale is the possible region for observed pileup in \fermi-GBM. The lower right panel compares the \Ep (i.e., Hardness) of the \fermi-GBM detected bursts. GRB 230812B is harder than the GRB 171010A and softer than the GRB 130427A at a similar redshift. GRB 230812B with \tninty $>$ 2 s lies around the dividing line of the population of SGRBs and LGRBs. However, based on the observed underlying SN and other observed properties, GRB 230812B is classified as {a} LGRB. Our analysis indicates that GRB 230812B is one of the brightest \fermi-GBM and LAT-detected LGRB {with the hardest} prompt spectra.

\section{Summary and Conclusion}
This work presents a detailed prompt and afterglow analysis of \fermi-LAT GRB 230812B. Due to its extreme brightness, the detecting instruments (\fermi-GBM) suffer a pileup in the {time range} [0.5-1.4] s, which we exclude from our analysis. The time-integrated spectral analysis showed that the spectrum in the \fermi-GBM energy range is the best fit by the \sw{Band+BB} function. In addition to this, all the time-resolved spectra favour an additional thermal component along with a dominant {non-thermal} component. All {of these} findings indicate the hybrid (baryonic + magnetic) jet composition of the burst. Spectral parameter evolution from our time-resolved analysis showed that the low energy spectral index $\alpha$ crosses the slow and fast cooling limit imposed by the synchrotron emission mechanism. The evolution of $\alpha$ poses a challenge to linking the {non-thermal} prompt emission with pure synchrotron emission. In addition, both $\alpha$ and \Ep were found to track the intensity of the burst throughout the prompt emission duration. {The} X-ray and optical LCs of this burst do not show any signature of late central engine activity (flare or plateau). Both X-ray and optical emissions are consistent with the closure relation in spectral regime $\nu_{m}$ $<$ $\nu_{o}$ $<$ $\nu_{x}$ $<$ $\nu_{c}$ in the ISM-like surrounding medium. {The} \fermi-LAT analysis showed that in the early stage ($<$ 10 ks), the LAT spectral index $\beta_{\rm LAT}$ is harder than the $\beta_{\rm x}$, indicating {that the} LAT emission is coming from a different spectral regime or entirely different origin. With the assumption of synchrotron origin {of the LAT photons}, we calculated that the break frequency $\nu_{c}$ may lie or remain between the X-ray and LAT frequencies throughout the {observations}. However, the synchrotron emission mechanism is unable to explain the origin of the 72 GeV {photons} detected by LAT at \tzero + 32s $>>$ \tninty, i.e., during the afterglow phase. The most plausible explanation for the VHE-detected photons is IC or SSC boosting of prompt or early afterglow photons by the shock-accelerated electrons in the external medium. The comparison of X-ray and optical LCs of GRB 230812B with similar bursts (GRB 130427A and 171010A) revealed that, unlike the bright prompt emission, the afterglow of GRB 230812B was fainter than the other SN-detected {GRBs} at a similar \Ep, E$_{\rm \gamma, iso}$, and redshift. Moreover, previous studies have shown that low-redshift GeV-TeV detected GRBs, such as GRB 130427A and GRB 201015A, are associated with supernovae \citep{2023ApJ...942...34R}. This event further reinforces the idea that the association of low-redshift GeV-TeV detected GRBs with underlying supernovae is a common occurrence.\\

\textbf{Acknowledgments:} AKR are thankful to Dr. Rahul Gupta for his continuous support during the analysis and writing of the paper. SBP acknowledges the financial support of ISRO under the AstroSat archival data utilization program (DS$\_$2B-13013(2)/1/2021-Sec.2). SBP also acknowledges support from DST grant no. DST/ICD/BRICS/Call-5/CoNMuTraMO/2023(G) for the present work. AA acknowledges funds and assistance provided by the Council of Scientific \& Industrial Research (CSIR), India, under file no. 09/948(0003)/2020-EMR-I. AA also acknowledges the Yushan Young Fellow Program by the Ministry of Education, Taiwan, for financial support. AJCT acknowledges support from the Spanish Ministry project PID2020-118491GB-I00 and Junta de Andalucia grant P20\_010168. This research has used data obtained through the HEASARC Online Service, provided by the NASA-GSFC, in support of NASA High Energy Astrophysics Programs.


\begin{thebibliography}
\bibitem[Kumar \& Zhang(2015)]{2015PhR...561....1K} Kumar, P. \& Zhang, B.\ 2015, \physrep, 561, 1. doi:10.1016/j.physrep.2014.09.008
\bibitem[Srinivasaragavan et al.(2024)]{2024ApJ...960L..18S} Srinivasaragavan, G.~P., Swain, V., O'Connor, B., et al.\ 2024, \apjl, 960, L18. doi:10.3847/2041-8213/ad16e7
\bibitem[Hussenot-Desenonges et al.(2023)]{2023arXiv231014310H} Hussenot-Desenonges, T., Wouters, T., Guessoum, N., et al.\ 2023, arXiv:2310.14310. doi:10.48550/arXiv.2310.14310
\bibitem[Beardmore et al.(2023)]{2023GCN.34400....1B} Beardmore, A.~P., Melandri, A., Sbarrato, T., et al.\ 2023, GRB Coordinates Network, Circular Service, No. 34400, 34400
\bibitem[Casentini et al.(2023)]{2023GCN.34402....1C} Casentini, C., Longo, F., Pittori, C., et al.\ 2023, GRB Coordinates Network, Circular Service, No. 34402, 34402
\bibitem[de Ugarte Postigo et al.(2023)]{2023GCN.34409....1D} de Ugarte Postigo, A., Agui Fernandez, J.~F., Thoene, C.~C., et al.\ 2023, GRB Coordinates Network, Circular Service, No. 34409, 34409
\bibitem[Evans \& Swift Team(2023)]{2023GCN.34388....1E} Evans, P.~A. \& Swift Team\ 2023, GRB Coordinates Network, Circular Service, No. 34388, 34388
\bibitem[Fermi GBM Team(2023)]{2023GCN.34386....1F} Fermi GBM Team\ 2023, GRB Coordinates Network, Circular Service, No. 34386, 34386
\bibitem[Frederiks et al.(2023)]{2023GCN.34403....1F} Frederiks, D., Lysenko, A., Ridnaia, A., et al.\ 2023, GRB Coordinates Network, Circular Service, No. 34403, 34403
\bibitem[Kennea \& Swift Team(2023)]{2023GCN.34393....1K} Kennea, J.~A. \& Swift Team\ 2023, GRB Coordinates Network, Circular Service, No. 34393, 34393
\bibitem[Lesage et al.(2023)]{2023GCN.34387....1L} Lesage, S., Burns, E., Dalessi, S., et al.\ 2023, GRB Coordinates Network, Circular Service, No. 34387, 34387
\bibitem[Page \& Swift-XRT Team(2023)]{2023GCN.34394....1P} Page, K.~L. \& Swift-XRT Team\ 2023, GRB Coordinates Network, Circular Service, No. 34394, 34394
\bibitem[Roberts et al.(2023)]{2023GCN.34391....1R} Roberts, O.~J., Meegan, C., Lesage, S., et al.\ 2023, GRB Coordinates Network, Circular Service, No. 34391, 34391
\bibitem[Scotton et al.(2023)]{2023GCN.34392....1S} Scotton, L., Kocevski, D., Racusin, J., et al.\ 2023, GRB Coordinates Network, Circular Service, No. 34392, 34392
\bibitem[Xiong et al.(2023)]{2023GCN.34401....1X} Xiong, S., Liu, J., Huang, Y., et al.\ 2023, GRB Coordinates Network, Circular Service, No. 34401, 34401
\bibitem[Roberts et al.(2023)]{2023GCN.34694....1R} Roberts, O.~J., Lesage, S., Cleveland, W., et al.\ 2023, GRB Coordinates Network, Circular Service, No. 34694, 34694
\bibitem[Kouveliotou et al.(1993)]{1993ApJ...413L.101K} Kouveliotou, C., Meegan, C.~A., Fishman, G.~J., et al.\ 1993, \apjl, 413, L101. doi:10.1086/186969
\bibitem[Meegan et al.(2009)]{2009ApJ...702..791M} Meegan, C., Lichti, G., Bhat, P.~N., et al.\ 2009, \apj, 702, 791. doi:10.1088/0004-637X/702/1/791
\bibitem[Amati(2006)]{2006MNRAS.372..233A} Amati, L.\ 2006, \mnras, 372, 233. doi:10.1111/j.1365-2966.2006.10840.x
\bibitem[Hjorth et al.(2003)]{2003Natur.423..847H} Hjorth, J., Sollerman, J., M{\o}ller, P., et al.\ 2003, \nat, 423, 847. doi:10.1038/nature01750
\bibitem[Ahumada et al.(2021)]{2021NatAs...5..917A} Ahumada, T., Singer, L.~P., Anand, S., et al.\ 2021, Nature Astronomy, 5, 917. doi:10.1038/s41550-021-01428-7
\bibitem[Troja et al.(2022)]{2022Natur.612..228T} Troja, E., Fryer, C.~L., O'Connor, B., et al.\ 2022, \nat, 612, 228. doi:10.1038/s41586-022-05327-3
\bibitem[Levan et al.(2023)]{2023arXiv230702098L} Levan, A., Gompertz, B.~P., Salafia, O.~S., et al.\ 2023, arXiv:2307.02098. doi:10.48550/arXiv.2307.02098
\bibitem[de Ugarte Postigo et al.(2011)]{2011A&A...525A.109D} de Ugarte Postigo, A., Horv{\'a}th, I., Veres, P., et al.\ 2011, \aap, 525, A109. doi:10.1051/0004-6361/201015261
\bibitem[Bo{\"e}r et al.(2015)]{2015ApJ...800...16B} Bo{\"e}r, M., Gendre, B., \& Stratta, G.\ 2015, \apj, 800, 16. doi:10.1088/0004-637X/800/1/16
\bibitem[Band et al.(1993)]{1993ApJ...413..281B} Band, D., Matteson, J., Ford, L., et al.\ 1993, \apj, 413, 281. doi:10.1086/172995
\bibitem[Kaneko et al.(2006)]{2006ApJS..166..298K} Kaneko, Y., Preece, R.~D., Briggs, M.~S., et al.\ 2006, \apjs, 166, 298. doi:10.1086/505911
\bibitem[Wang et al.(2024)]{2024RAA....24b5006W} Wang, W.-K., Xie, W., Gao, Z.-F., et al.\ 2024, Research in Astronomy and Astrophysics, 24, 025006. doi:10.1088/1674-4527/ad16af
\bibitem[Burgess et al.(2020)]{2020NatAs...4..174B} Burgess, J.~M., B{\'e}gu{\'e}, D., Greiner, J., et al.\ 2020, Nature Astronomy, 4, 174. doi:10.1038/s41550-019-0911-z
\bibitem[Norris et al.(1986)]{1986ApJ...301..213N} Norris, J.~P., Share, G.~H., Messina, D.~C., et al.\ 1986, \apj, 301, 213. doi:10.1086/163889
\bibitem[Golenetskii et al.(1983)]{1983Natur.306..451G} Golenetskii, S.~V., Mazets, E.~P., Aptekar, R.~L., et al.\ 1983, \nat, 306, 451. doi:10.1038/306451a0
\bibitem[Li et al.(2021)]{2021ApJS..254...35L} Li, L., Ryde, F., Pe'er, A., et al.\ 2021, \apjs, 254, 35. doi:10.3847/1538-4365/abee2a
\bibitem[Ror et al.(2023)]{2023ApJ...942...34R} Ror, A.~K., Gupta, R., Jel{\'\i}nek, M., et al.\ 2023, \apj, 942, 34. doi:10.3847/1538-4357/aca414
\bibitem[Ror et al.(2024)]{2024ApJ...971..163R} Ror, A.~K., Gupta, R., Aryan, A., et al.\ 2024, \apj, 971, 163. doi:10.3847/1538-4357/ad5554
\bibitem[Ror et al.(2024)]{2024BSRSL..93..709R} Ror, A.~K., Bhushan Pandey, S., Gupta, R., et al.\ 2024, Bulletin de la Societe Royale des Sciences de Liege, 93, 709. doi:10.25518/0037-9565.11848
\bibitem[Gupta et al.(2021)]{2021MNRAS.505.4086G} Gupta, R., Oates, S.~R., Pandey, S.~B., et al.\ 2021, \mnras, 505, 4086. doi:10.1093/mnras/stab1573
\bibitem[Gupta et al.(2021)]{2021RMxAC..53..113G} Gupta, R., Pandey, S.~B., Castro-Tirado, A.~J., et al.\ 2021, Revista Mexicana de Astronomia y Astrofisica Conference Series, 53, 113. doi:10.22201/ia.14052059p.2021.53.23
\bibitem[Gupta et al.(2022)]{2022MNRAS.511.1694G} Gupta, R., Gupta, S., Chattopadhyay, T., et al.\ 2022, \mnras, 511, 1694. doi:10.1093/mnras/stac015
\bibitem[Gupta et al.(2022)]{2022JApA...43...11G} Gupta, R., Kumar, A., Pandey, S.~B., et al.\ 2022, Journal of Astrophysics and Astronomy, 43, 11. doi:10.1007/s12036-021-09794-4
\bibitem[Gupta(2023)]{2023arXiv231216265G} Gupta, R.\ 2023, arXiv:2312.16265. doi:10.48550/arXiv.2312.16265
\bibitem[Gupta et al.(2024)]{2024ApJ...972..166G} Gupta, R., Pandey, S.~B., Gupta, S., et al.\ 2024, \apj, 972, 166. doi:10.3847/1538-4357/ad5a92
\bibitem[Castro-Tirado et al.(2024)]{2024A&A...683A..55C} Castro-Tirado, A.~J., Gupta, R., Pandey, S.~B., et al.\ 2024, \aap, 683, A55. doi:10.1051/0004-6361/202346042
\bibitem[Caballero-Garc{\'\i}a et al.(2023)]{2023MNRAS.519.3201C} Caballero-Garc{\'\i}a, M.~D., Gupta, R., Pandey, S.~B., et al.\ 2023, \mnras, 519, 3201. doi:10.1093/mnras/stac3629
\bibitem[Vianello et al.(2015)]{2015arXiv150708343V} Vianello, G., Lauer, R.~J., Younk, P., et al.\ 2015, arXiv:1507.08343. doi:10.48550/arXiv.1507.08343
\bibitem[Burgess(2014)]{2014MNRAS.445.2589B} Burgess, J.~M.\ 2014, \mnras, 445, 2589. doi:10.1093/mnras/stu1925
\bibitem[Vianello et al.(2018)]{2018ApJ...864..163V} Vianello, G., Gill, R., Granot, J., et al.\ 2018, \apj, 864, 163. doi:10.3847/1538-4357/aad6ea
\bibitem[Barthelmy et al.(2005)]{2005SSRv..120..143B} Barthelmy, S.~D., Barbier, L.~M., Cummings, J.~R., et al.\ 2005, \ssr, 120, 143. doi:10.1007/s11214-005-5096-3
\bibitem[Evans et al.(2007)]{2007A&A...469..379E} Evans, P.~A., Beardmore, A.~P., Page, K.~L., et al.\ 2007, \aap, 469, 379. doi:10.1051/0004-6361:20077530
\bibitem[Evans et al.(2009)]{2009MNRAS.397.1177E} Evans, P.~A., Beardmore, A.~P., Page, K.~L., et al.\ 2009, \mnras, 397, 1177. doi:10.1111/j.1365-2966.2009.14913.x
\bibitem[Fraija et al.(2019)]{2019ApJ...879L..26F} Fraija, N., Dichiara, S., Pedreira, A.~C.~C. do E.~S., et al.\ 2019, \apjl, 879, L26. doi:10.3847/2041-8213/ab2ae4
\bibitem[MAGIC Collaboration et al.(2019)]{2019Natur.575..459M} MAGIC Collaboration, Acciari, V.~A., Ansoldi, S., et al.\ 2019, \nat, 575, 459. doi:10.1038/s41586-019-1754-6
\bibitem[Fraija et al.(2021)]{2021ApJ...918...12F} Fraija, N., Veres, P., Beniamini, P., et al.\ 2021, \apj, 918, 12. doi:10.3847/1538-4357/ac0aed
\bibitem[H.~E.~S.~S. Collaboration et al.(2021)]{2021Sci...372.1081H} H.~E.~S.~S. Collaboration, Abdalla, H., Aharonian, F., et al.\ 2021, Science, 372, 1081. doi:10.1126/science.abe8560
\bibitem[Arnaud(1996)]{1996ASPC..101...17A} Arnaud, K.~A.\ 1996, Astronomical Data Analysis Software and Systems V, 101, 17
\bibitem[Galama et al.(1998)]{1998Natur.395..670G} Galama, T.~J., Vreeswijk, P.~M., van Paradijs, J., et al.\ 1998, \nat, 395, 670. doi:10.1038/27150
\bibitem[Liang et al.(2010)]{2010ApJ...725.2209L} Liang, E.-W., Yi, S.-X., Zhang, J., et al.\ 2010, \apj, 725, 2209. doi:10.1088/0004-637X/725/2/2209
\bibitem[Sari et al.(1998)]{1998ApJ...497L..17S} Sari, R., Piran, T., \& Narayan, R.\ 1998, \apjl, 497, L17. doi:10.1086/311269
\bibitem[Abbott et al.(2017)]{2017ApJ...848L..13A} Abbott, B.~P., Abbott, R., Abbott, T.~D., et al.\ 2017, \apjl, 848, L13. doi:10.3847/2041-8213/aa920c
\bibitem[Pe'er(2015)]{2015AdAst2015E..22P} Pe'er, A.\ 2015, Advances in Astronomy, 2015, 907321. doi:10.1155/2015/907321
\bibitem[Zhang et al.(2018)]{2018NatAs...2...69Z} Zhang, B.-B., Zhang, B., Castro-Tirado, A.~J., et al.\ 2018, Nature Astronomy, 2, 69. doi:10.1038/s41550-017-0309-8
\bibitem[Preece et al.(2000)]{2000ApJS..126...19P} Preece, R.~D., Briggs, M.~S., Mallozzi, R.~S., et al.\ 2000, \apjs, 126, 19. doi:10.1086/313289
\bibitem[Kumar et al.(2022)]{2022NewA...9701889K} Kumar, A., Pandey, S.~B., Gupta, R., et al.\ 2022, \na, 97, 101889. doi:10.1016/j.newast.2022.101889
\bibitem[Xu et al.(2013)]{2013ApJ...776...98X} Xu, D., de Ugarte Postigo, A., Leloudas, G., et al.\ 2013, \apj, 776, 98. doi:10.1088/0004-637X/776/2/98
\bibitem[Goldstein et al.(2023)]{gbmdatatool} Adam Goldstein, William H. Cleveland et al. \ 2023
\bibitem[Scargle et al.(2013)]{2013ApJ...764..167S} Scargle, J.~D., Norris, J.~P., Jackson, B., et al.\ 2013, \apj, 764, 167. doi:10.1088/0004-637X/764/2/167
\bibitem[Mukherjee et al.(1998)]{1998ApJ...508..314M} Mukherjee, S., Feigelson, E.~D., Jogesh Babu, G., et al.\ 1998, \apj, 508, 314. doi:10.1086/306386
\bibitem[Ackermann et al.(2014)]{2014Sci...343...42A} Ackermann, M., Ajello, M., Asano, K., et al. \ 2014, Science, 343, 42. doi:10.1126/science.1242353
\bibitem[Pandey et al.(2021)]{2021MNRAS.507.1229P} Pandey, S.~B., Kumar, A., Kumar, B., et al.\ 2021, \mnras, 507, 1229. doi:10.1093/mnras/stab1889
\bibitem[Wang et al.(2024)]{wang2024GRB230812B} Wang, C.-Y., Yin, Y.-H. I., Zhang, B.-B., et al. 2024, arXiv:2409.12613

\end{thebibliography}
\end{document}